\documentclass[twoside,12pt]{article}
\usepackage{epsfig}

\def\Journal#1#2#3#4{{#1} {#2} (#4) #3 }

\def\NPA{{\em Nucl. Phys.} A}

\def\PLB{{\em Phys. Lett.} B}

\def\PRL{\em Phys. Rev. Lett.}

\def\PREP{\em Phys. Rep.}

\def\PRD{{\em Phys. Rev.} D}
\def\PRC{{\em Phys. Rev.} C}

\newcommand{\be}{\begin{equation}}
\newcommand{\ee}{\end{equation}}
\newcommand{\bea}{\begin{eqnarray}}
\newcommand{\eea}{\end{eqnarray}}

\begin{document}

\title{ \vspace{1cm} Open charm hadron spectroscopy at B-factories}
\author{Y.\ Kato$^1$ and T.\ Iijima$^{1,2}$ 
\\
$^1$Kobayashi-Maskawa Institute, Nagoya University, Nagoya 464-8602, Japan \\
$^2$Graduate School of Science, Nagoya University, Nagoya 464-8602, Japan \\
}

\maketitle
\begin{abstract} 

Open charm hadrons are excellent probes to study the dynamics of quarks and gluons
inside hadrons. Because the mass of a charm quark is heavier than $\Lambda_{QCD}$, 
so called heavy quark symmetry emerges in the hadron containing a charm quark and plays a central role in the
classification and constraining the interactions.
Belle and BaBar, which are B-factory experiments, have made significant advances
in the field of open charm hadron spectroscopy in these 15 years. In this article, experimental 
advances on both open charm mesons and baryons by B-factory experiments are reviewed. 
Finally, the prospect of the open charm hadrons with the next generation B-factory experiment, Belle II is described. 

\end{abstract}

\section{Introduction} \label{sec_intro}
Isolated quarks or gluons have never been observed. 
Only their bound state known as hadrons have been observed. Hadron dynamics 
is governed by quantum chromodynamics (QCD), which is non-perturvative in the energy scale of a hadron. 
Consequently, a couple of theoretical methods and models have been developed to describe them. 
These include the quark model, lattice QCD, QCD sum rule, and so forth. In particular, despite its relatively simple form, the quark model 
has reproduced the spectra of experimentally observed hadrons fairly well~\cite{Isgur:1991wq,Isgur:1979be,Isgur:1977ef,Isgur:1978wd} 
with some exceptions like $\Lambda(1405)$ and Roper. Therefore, the quark model has been the basis of 
our understanding of hadrons. However, the connection between QCD and quark model remains unclear from first principles.
To better understand the dynamics inside hadrons, open flavor heavy hadrons, which contain one charm or bottom quark as constituent, is a good probe.

In the limit of $m_{Q} \to \infty$, two symmetries appear in QCD~\cite{Isgur:1991wq}. One is the heavy quark flavor symmetry,
where the dynamics becomes independent of the flavor of heavy quarks. The other one is heavy quark spin symmetry (HQSS),
where the spin of a heavy quark becomes a conserved value. The $SU(2)$ symmetry from HQSS and $U(N_{h})$,
where $N_{h}$ is the number of heavy flavor quarks, from heavy quark flavor symmetry lead to the
$U(2N_{h})$ classification scheme for the open heavy hadrons.
The mass of a charm quark is much larger than $\Lambda_{QCD}$, which is the energy scale of QCD. 
Therefore, these two symmetries are approximately applicable to open charm hadrons. 
Section ~\ref{sec_hqss} describes details of the classification scheme by HQSS.

Besides these theoretical interests, open charm hadrons also have an experimental advantage.
They have narrow widths compared to hadrons without heavy quarks. For example, in the meson sector, the $J^{P} = 1^{-}$ state
$\rho$ is composed of up and down quarks. It has a width of more than 100 MeV, whereas the corresponding $1^{-}$ state $D^{\ast}$ has a 
width less than $O(1)$ MeV. The $1/2^{-}$ and $3/2^{-}$ states in the $\Lambda$ baryons,
$\Lambda (1405)$ and $\Lambda(1520)$ have widths of about 50 MeV and 15 MeV, respectively,
while the corresponding states in the charm sector, $\Lambda_{c}(2593)^{+}$
and $\Lambda_{c}(2625)^{+}$ have widths of around 2.5 MeV and $<1.0$ MeV, respectively. 
Thanks to the narrow width, it is experimentally easy to discover open charm hadrons and measure various properties.

This paper reviews recent experimental progress on open charm hadrons with an emphasis on the roles played by B-factory experiments, Belle and BaBar.
There have been revolutionary advances on  heavy quark hadrons (not only open heavy hadrons, but also quarkonium states) 
spectroscopy since Belle and BaBar have begun collecting data. 
A huge number of hadrons that cannot be understood as ordinary mesons have been discovered,
including charmonium-like states such as $X(3872)$~\cite{Choi:2003ue}, $Y(4260)$~\cite{Aubert:2005rm},
and $Z(4430)$~\cite{Choi:2007wga}, which are generically called ``$XYZ$" states, and bottomonium-like states
$Z_{b}(10610)$ and $Z_{b}(10650)$~\cite{Belle:2011aa}.
A review of exotic quarkonium-like hadrons driven by B-factory experiments can be found in~\cite{Hosaka:2016pey}.
Additionally, conventional charmonium and bottomonium states such as $\eta_{c}(2S)$ or $\eta_{b}(2S)$ have been discovered~\cite{Choi:2002na,Mizuk:2012pb}.
Finally, more than fifteen new open charm hadrons have been discovered and their properties comprehensively measured. 
This review focuses on the studies of these newly discovered open charm hadrons.

\section{B-factory experiments} \label{sec_exp}
The Belle experiment at KEK in Japan and the BaBar experiment at SLAC in the US are designed and optimized 
to test Kobayashi-Maskawa's theory for the $CP$ violation~\cite{Kobayashi:1973fv} by 
studying $B$ meson decays into the $CP$ eigenstates such as $J/\psi K^{0}_{S}$.  
Two accelerators, KEKB~\cite{Kurokawa:2001nw,Abe:2013kxa} and PEP-II are 
asymmetric energy $e^{+} e^{-}$ colliders operated mainly at a center-of-mass energy of 10.58 GeV.
This energy corresponds to the mass of an excited bottomonium state $\Upsilon(4S)$, which decays into a $B$ meson pair
with a branching fraction of almost 100$\%$, and is the origin of the name B-factory. 
The Belle~\cite{Abashian:2000cg} and BaBar~\cite{Aubert:2001tu} detectors are designed as 4$\pi$ general
purpose spectrometer in a solenoidal magnetic field.
These detectors have some common features such as a good vertexing capability, charged particle identification, 
and photon energy measurement by the electromagnetic calorimeter to reconstruct various B-meson decays. 
The number of $B$ meson pairs recorded are $772 \times 10^{6}$ for Belle and $466.5 \times 10^{6}$ for BaBar.  
The accumulated integrated luminosity is 980 fb$^{-1}$ for Belle and 468 fb$^{-1}$ for BaBar, this includes  
data with a center-of-mass energy other than $\Upsilon(4S)$.
The sizes of data samples collected by B-factory experiments are more than two orders of magnitude higher than that accumulated by older experiments.
The high statistics data and general purpose feature of B-factory experiments allow to study not only
the $CP$ violation in $B$ decay but also rare $B$ decays, $\tau$ and charm physics, and hadron spectroscopy. 
The data with $B$ meson pairs are used to study open charm hadrons produced via $B$ meson decay, while
the data sample including other center-of-mass energies are used for those produced via the $e^{+}e^{-} \to c\bar{c}$ reaction.
Section~\ref{sec_prod} provides additional details about the production mechanism.

\section{Production of open charm hadrons in B-factory experiments} \label{sec_prod}
Open charm hadrons are produced mainly in two processes at B-factory experiments.
One is decay of the $B$ meson and the other is the $e^{+} e^{-} \to c\bar{c}$ 
reaction followed by hadronization of the (anti) charm quark (also called continuum production).
Each process has advantages and disadvantages in terms of charm hadron spectroscopy.
They serve complementary roles. 

In the $B$ decays, charmed hadrons are produced mainly via the Cabibbo favored $b \to c$ transition.
In almost of all cases, not only charmed hadrons but also $B$ mesons are reconstructed exclusively.
This leads to a good signal purity compared with the $e^{+} e^{-} \to c\bar{c}$ reaction. 
For a two body decay into charmed hadrons and known long-lived hadrons (such as pions or protons),
the helicity of the charmed hadrons is constrained as the spin of $B$ meson is zero. 
This leads to a model-independent determination of the spin of charmed hadrons. 
Additionally, as the final state is clearly defined after reconstructing the $B$ meson, the parity 
can be also determined from the interference with the non-resonant decay processes.
However, the production of the high spin state is suppressed and cannot be studied in $B$ meson decays.   

In the $e^{+} e^{-} \to c\bar{c}$ reaction, only a charmed hadron is reconstructed exclusively in almost all cases.
Therefore, the background level is relatively high.
To reduce the background, the momentum of a charmed hadron in the center-of-mass frame is required to be high 
as the main background source is the combinatoric one, which tends to have a lower momentum. 
The determination of the spin is not trivial because there are no constraint on the helicity of charmed hadron from the production,
and it should be determined by the fit to the decay angular distribution.
On the other hand, the production rate is relatively higher than that from $B$ decay and  
higher spin states are not suppressed. Therefore, the number of signal events are higher than
that from $B$ meson decays. Actually, the number of open charm hadrons discovered in B-factories 
via the $e^{+} e^{-} \to c\bar{c}$ is larger than that from $B$ decays for both mesons and baryons.

\section{Classification of open charm hadrons with the heavy quark spin symmetry} \label{sec_hqss}
The strength of the chromo-magnetic interaction is proportional to the inverse of the product of masses of two quarks.
Therefore, in the limit of heavy quark mass $m_{Q} \to \infty$, the magnetic interaction disappears and 
the spin of the heavy quark $S_{Q}$ becomes a conserved quantum number.
As the total spin $J=S_{Q}+j_{q}$, where $j_{q}$ is the sum of the light quark spin and orbital
angular momentum $L$, is also a conserved, $j_{q}$ is also a good quantum number. 
Open heavy hadrons with $J=j_{q} \pm 1/2$ are called a heavy quark spin doublet, 
which degenerate at the limit of $m_{Q} \to \infty$. 
This classification scheme is very useful to understand the spectrum of open heavy hadrons.

The simplest example of a heavy quark doublet in the $c\bar{q}$ meson, where
$q$ is $u$ or $d$, is the ground state $D$ and $D^{\ast}$ mesons. In this case, $j_{q}=1/2$ and total spin 
are 0 and 1. Because a charm quark has a finite mass, the HQSS is broken in the charmed hadron,
and $D$ and $D^{\ast}$ have mass difference of around 140 MeV/$c^{2}$. This is much smaller than those of 
$\pi$ and $\rho$ (about 630 MeV/$c^{2}$) and $K$ and $K^{\ast}$ (about 400 MeV/$c^{2}$). 
An example in the baryon sector is $\Sigma_{c}(2455)$ and $\Sigma_{c}(2520)$, which has $j_{q}=1$ and
the total spins are $1/2$ and $3/2$. The mass difference is around 65 MeV/$c^{2}$, which is
much smaller than that for $\Sigma$ and $\Sigma(1385)$ (around 190 MeV/$c^{2}$).
In the quark model, the first excited state of both mesons and baryons should have $L=1$. 
In case of the $D$ meson family, $j_{q}$ has two values: $1 \pm 1/2 = 1/2,3/2$. Therefore, there are two 
heavy quark spin doublets. They are identified as $D_{1} (2420)$, $D_{2}^{\ast}(2460)$ for $j_{q}=3/2$ and 
$D_{0}^{\ast} (2400)$, $D_{1}(2430)$ for $j_{q}=1/2$.  
When a new state is observed, the concept of a heavy quark spin doublet is the guiding principle 
to understand the nature of the observed state. 

It is worth mentioning that combination of heavy quark symmetry and chiral symmetry,
which is an approximated symmetry in light quark sector in QCD, is very useful to describe
the interactions of heavy mesons such as one-pion transition~\cite{Wise:1992hn,Kilian:1992hq}.


\section{Spectroscopy of charmed mesons}

\subsection{$D$ meson family} 
\begin{figure*}[htbp]
  \begin{center}
    \includegraphics[scale=0.5]{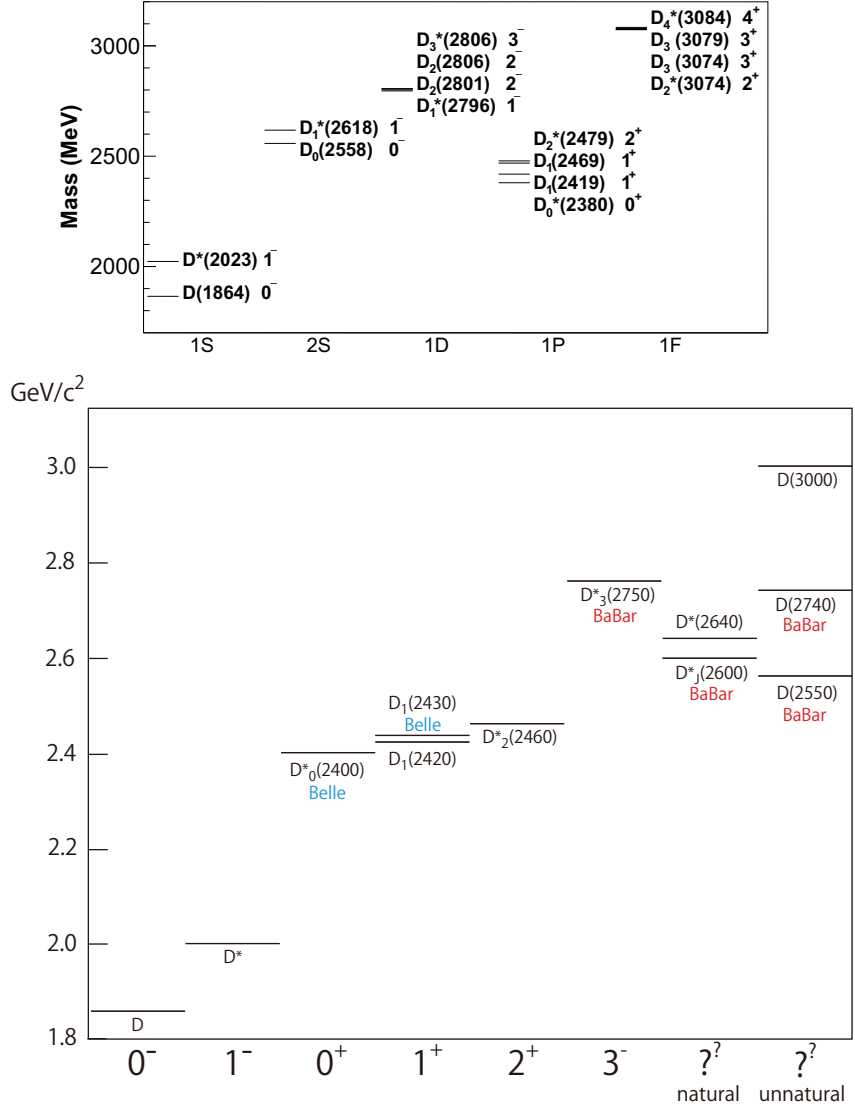}
    \caption{Top:Theoretically expected $D$ meson family from~\cite{Aaij:2013sza} based on the
             modified Godfrey-Isgur prediction~\cite{Godfrey:1985xj}. Bottom: Currently observed $D$ meson family. 
	     For the label of hadrons, convention of Ref.~\cite{Patrignani:2016xqp} is followed.
             Name of the experiment indicates the first observation occurred in Belle or BaBar.}
    \label{charmmeson_spectrum}
  \end{center}
\end{figure*}

The member of $D$ mesons family are composed of $c$ quark and $\bar{u}$ or $\bar{d}$ quark.
Before the start of B-factory experiments, four states were observed among the $D$ meson family.
Two were the ground state $D$ and $D^{\ast}$, which are the heavy quark spin doublet $j_{q}=1/2$ with negative parity. 
The other two were the heavy quark spin doublet with $L=1$ and $j_{q}=3/2$: $D_{1}(2420)$ and $D_{2}^{\ast}(2460)$. 
We label the excited $D$ meson family as $D_{J}^{(\ast)}$, where $J$ is the total spin. The asterisk denotes a
state with a natural parity ($0^{+}, 1^{-}, 2^{+}, 3^{-}...$).
For the states with $j_{q}=3/2$, the two-body decay with the ground state open charm meson
and $\pi$ must be in the $D$-wave to conserve $j_{q}$. Therefore, the width should be narrow and relatively easy to observe.
In contrast, for the state with $j_{q}=1/2$ and $L=1$, the same two-body decay should proceed in S-wave, and widths should be wide.
Therefore, these states are difficult to be observed experimentally and were not observed prior to the B-factory era. 
Figure \ref{charmmeson_spectrum} shows the theoretically or experimentally expected/observed members of the $D$ meson family.
For the states firstly observed by B-factory experiments, the name of experiment is included in the figure. 
It is obvious that B-factory experiments have made significant contribution on the spectroscopy of the $D$ meson family.
Table~\ref{summary_dfamily} summarizes masses, widths, decay modes of $D$ meson family. 

\begin{center}
  \begin{table*}[htbp]
	  \caption{Summary of the currently observed members of $D$ meson family~\cite{Patrignani:2016xqp}. 
		   The $J^{P}$ of $D_{1}(2420)^{+}$ is not determined
	           but assumed to be the same as isospin partner $D_{1}(2420)^{0}$.}
     \begin{tabular}{c|ccccc} \hline \hline
       Particle                   & Mass (MeV/c$^{2})$    &Width (MeV)          & $J^{P}$  & Strong or EM decay modes                                                                                               \\ \hline  
       $D^{0}$                    & $1864.83\pm0.05$      &-                    & $0^{-}$  & -                                                                                                                      \\
       $D^{+}$                    & $1869.65\pm0.05$      &-                    & $0^{-}$  & -                                                                                                                      \\ \hline
                                                                                         
       $D^{\ast}(2007)^{0}$       & $2006.85\pm0.05$      &$<2.1$               & $1^{-}$  & $D^{0} \pi^{0},D^{0} \gamma$                                                                                           \\
       $D^{\ast}(2010)^{+}$       & $2010.26\pm0.05$      &$0.0834\pm0.0018$    & $1^{-}$  & $D^{0} \pi^{+}, D^{+} \pi^{0}, D^{+} \gamma$                                                                           \\ \hline
                                                                                         
       $D_{0}^{\ast} (2400)^{0}$  & $2318\pm29$           &$267\pm40$           & $0^{+}$  & $D^{+} \pi^{-}$                                                                                                        \\ 
       $D_{0}^{\ast} (2400)^{+}$  & $2351\pm7$            &$230\pm17$           & $0^{+}$  & $D^{0} \pi^{+}$                                                                                                        \\ \hline
                                                                                         
       $D_{1} (2420)^{0}$         & $2420.8\pm0.5$        &$31.7\pm2.5$         & $1^{+}$  & $D^{\ast +} \pi^{-}, D^{0} \pi^{+} \pi^{-}, D^{0} \rho^{0}, D^{0} f_{0}(500), D_{0}^{ \ast}(2400)^{+} \pi^{-}$         \\ 
       $D_{1} (2420)^{+}$         & $2423.2\pm2.4$        &$25\pm6$             & $1^{+}$  & $D^{\ast 0} \pi^{+}, D^{+} \pi^{+} \pi^{-}, D^{+} \rho^{0}, D^{+} f_{0}(500), D_{0}^{ \ast}(2400)^{0} \pi^{+}$         \\ \hline
                                                                                         
       $D_{1} (2430)^{0}$         & $2427\pm40$           &$384^{+130}_{-110}$  & $1^{+}$  & $D^{\ast +} \pi^{-}$                                                                                                   \\ \hline
                                                                                         
       $D_{2}^{\ast} (2460)^{0}$  & $2460.7\pm0.4$        &$47.5\pm1.1$         & $2^{+}$  & $D^{+} \pi^{-}, D^{\ast +} \pi^{-}$                                                                                    \\ 
       $D_{2}^{\ast} (2460)^{+}$  & $2465.4\pm1.3$        &$46.7\pm1.2$         & $2^{+}$  & $D^{0} \pi^{+}, D^{\ast 0} \pi^{+}$                                                                                    \\ \hline
                                                                                         
       $D_{0} (2550)^{0}$         & $2564\pm20$           &$135\pm17$           & $?^{?}$  & $D^{\ast +} \pi^{-}$                                                                                                   \\ \hline
                                                                                         
       $D_{J}^{\ast} (2600)^{+/0}$& $2623\pm12$           &$139\pm31$           & $?^{?}$  & $D^{+} \pi^{-}, D^{0} \pi^{+}, D^{\ast +} \pi^{-}$                                                                     \\ \hline
       $D_{J}^{\ast} (2640)^{+}$  & $2637\pm6$            &$<15$                & $?^{?}$  & $D^{\ast +} \pi^{+}  \pi^{-}$                                                                                          \\ \hline
                                                                                         
       $D_{J} (2740)^{0}$         & $2737.0\pm12$         &$73\pm28.0$          & $?^{?}$  & $D^{\ast +} \pi^{-}$                                                                                                   \\ \hline
                                                                                         
       $D_{3}^{\ast} (2750)$      & $2763.5\pm3.4$        &$66\pm5$             & $3^{-}$  & $D^{+} \pi^{-}, D^{0} \pi^{+}, D^{\ast +} \pi^{-}$                                                                     \\ \hline
                                                                                         
       $D_{J} (3000)^{0}$         & $3214\pm60$           &$186\pm80$           & $?^{?}$  & $D^{\ast +} \pi^{-}$                                                                                                   \\ \hline

    \end{tabular}    
    \label{summary_dfamily}
  \end{table*}
\end{center}



\subsubsection{Studies from $B$ decays}
\begin{figure*}[htbp]
  \begin{center}
    \includegraphics[scale=0.35]{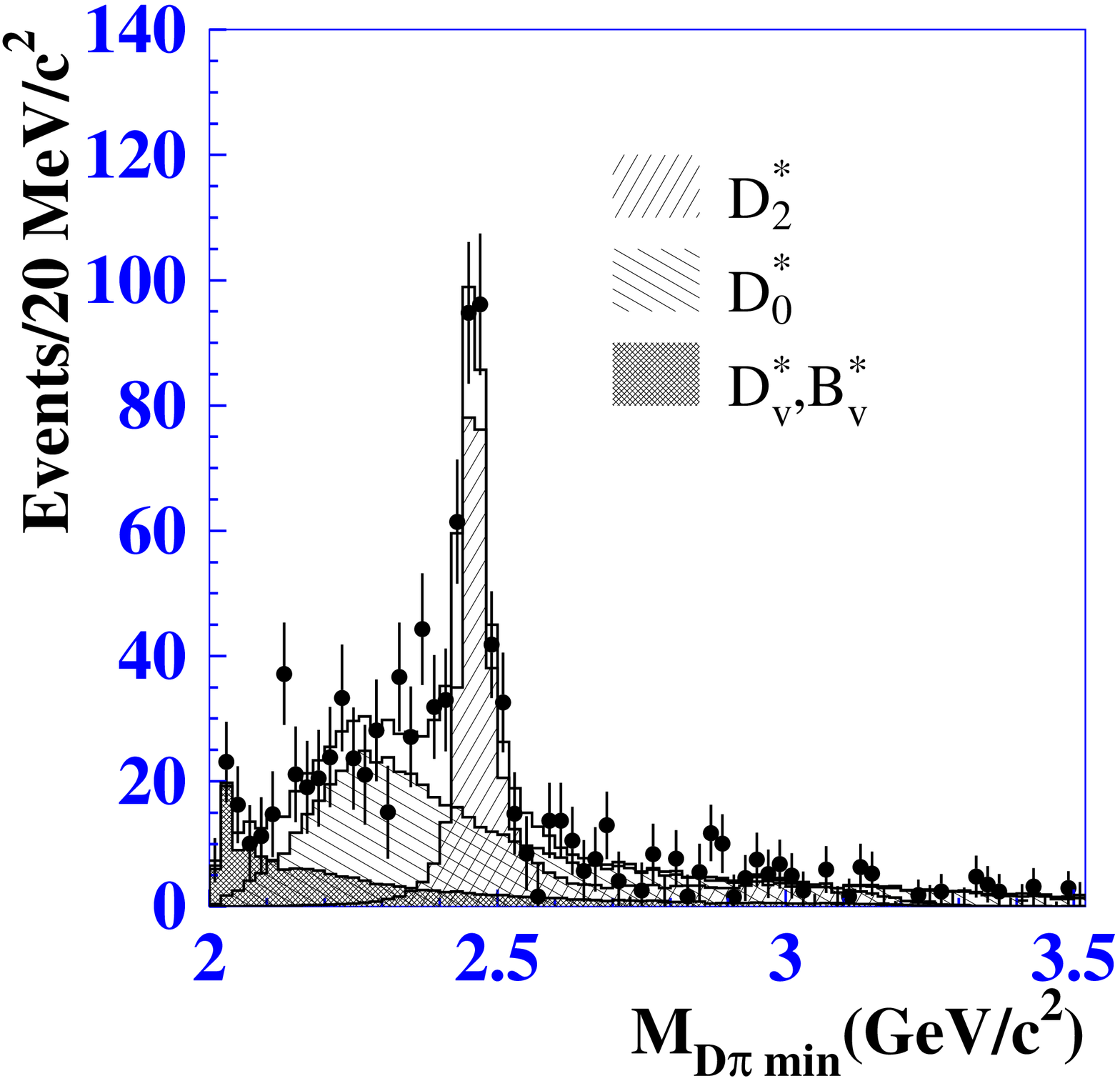}
    \includegraphics[scale=0.35]{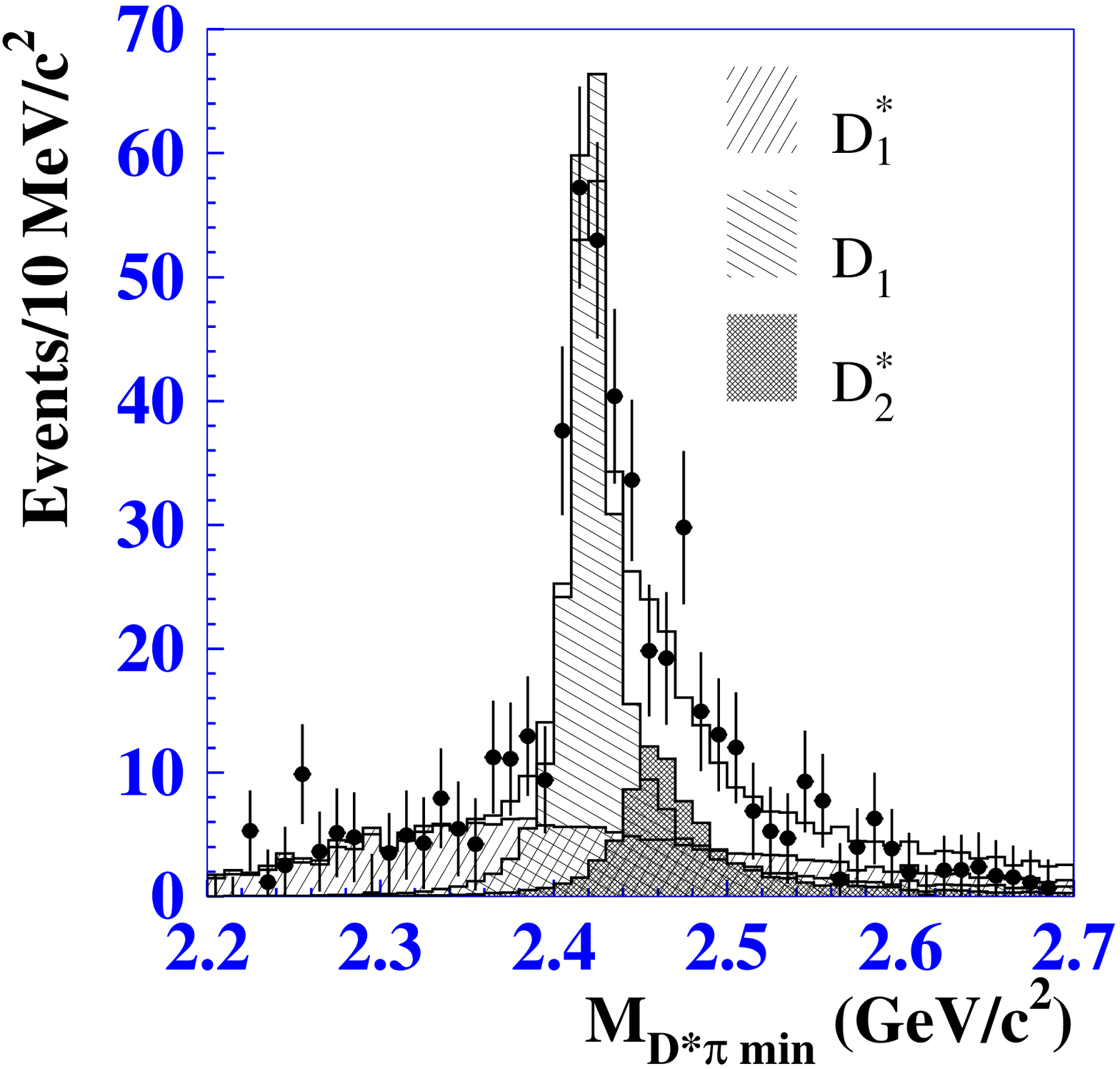}
    \caption{Minimal $D^{+} \pi^{-}$ (left) and $D^{\ast +} \pi^{-}$ (right) mass distribution for 
             $B^{-} \to D^{+} \pi^{-} \pi^{-}$ candidates showing the first observation of $D_{0}^{\ast}(2400)^{0}$
             from \cite{Abe:2003zm}.  }
    \label{dpi_belle}
  \end{center}
\end{figure*}

Belle studied excited $D$ mesons with Dalitz analysis in the decay $B^{-} \to D_{J}^{(\ast)} \pi^{-} \to  D^{(\ast) +} \pi^{-} \pi^{-}$
with a data sample of $65\times10^{6}$ $B\bar{B}$ pairs~\cite{Abe:2003zm}.
For the $B^{-} \to D^{+} \pi^{-} \pi^{-}$ decay, the fit was performed in the plane of 
the two $m^{2}(D^{0} \pi^{-})$.  Figure \ref{dpi_belle} (left) shows the minimal $D^{+} \pi^{-}$ distribution with the fitted result overlaid.
Besides a prominent peak corresponding to $D_{2}^{\ast}(2420)^{0}$, a bump structure corresponding to $D_{0}^{\ast}(2400)^{0}$ is observed. 
When $D_{0}^{\ast}(2400)^{0}$ is not included in the fit, or a state with $J^{P}$ other than $0^{+}$ is included, 
$-2ln(L)$ increases by more than 200. Thus, Belle claimed the first observation of $D_{0}^{\ast}(2400)^{0}$.  
The large width of $276\pm21\pm63$ MeV suggests that this decay proceeds in the S-wave. 
Figure \ref{dpi_belle} (right) shows the minimal $D^{\ast +} \pi^{-}$ distribution with fitted result overlaid for states
with common mixing angle $\omega$.  
In the fit, when the $D_{1}(2430)^{0}$ is not included, or a state with $J^{P}$ other than $1^{+}$ are included, the
$-2ln(L)$ increases by more than 100. Thus Belle claimed the first observation of $D_{1}(2430)^{0}$.  
To date, this is the only significant evidence for this state.
As the mass of the $c$ quark is finite and heavy quark spin symmetry is not perfect, there should be some 
mixing between the $S$ and $D$ waves. Therefore, $D_{1}(2420)$ and $D_{1}(2430)$ are expressed as mixing of the $S$ and $D$ wave.
The mixing angle is obtained as $\omega = -0.10\pm0.03\pm0.02\pm0.02$ rad, which means the $S$ and $D$ waves are dominant for 
$D_{1}(2420)$ and $D_{1}(2430)$, respectively.
BaBar also did the Dalitz analysis for $B^{-} \to D_{J}^{(\ast) 0} \pi^{-} \to  D^{+} \pi^{-} \pi^{-}$~\cite{Aubert:2009wg}
with a data sample around six times higher than Belle's analysis~\cite{Abe:2003zm}, confirming the existence of 
$D_{0}^{\ast}(2400)^{0}$ with a mass and width of $2297\pm8\pm20$ MeV/$c^{2}$ and $273\pm12\pm48$ MeV, respectively.
These values are consistent with Belle's result with better precision.

Belle also studied the charged partners in the $\bar{B}^{0} \to D_{J}^{\ast +} \pi^{+} \to D^{0} \pi^{+} \pi^{-}$ decays.  
The analysis method is basically the same as that for neutral states~\cite{Abe:2003zm}, where several resonances 
such as $\rho$, $f_{0}(600)$, $f_{0}(980)$, and $f_{0}(1380)$ are included in the $\pi^{+} \pi^{-}$ channel.
The mass and width of the $D_{0}^{\ast}(2400)^{+}$ were fixed with the values for the neutral partner. The significance of the 
$D_{0}^{\ast} (2400)^{+}$ is obtained as 6.8$\sigma$, which first observation of this state.

\subsubsection{Higher excited states in the continuum production}
\begin{figure*}[htbp]
  \begin{center}
    \includegraphics[scale=0.35]{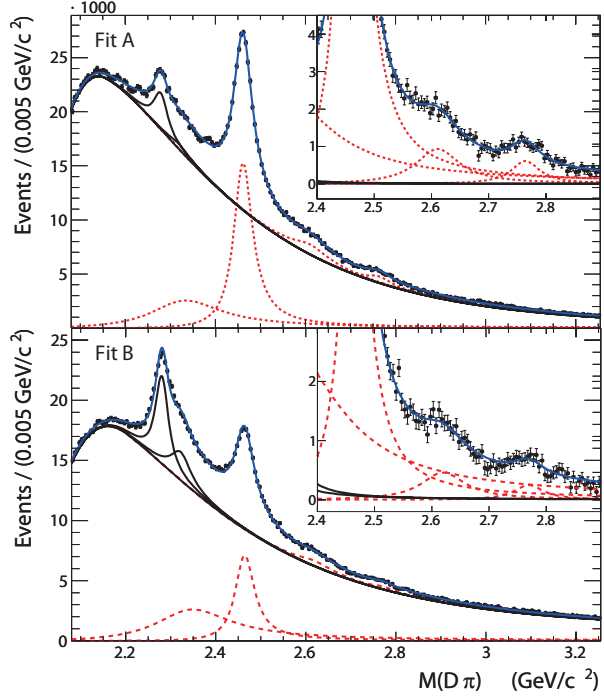}
    \caption{$D^{+} \pi^{-}$ (top) and $D^{0} \pi^{+}$ (left) invariant mass distributions
             from~\cite{delAmoSanchez:2010vq}.}
    \label{dpi_continuum}
  \end{center}
\end{figure*}

\begin{figure*}[htbp]
  \begin{center}
    \includegraphics[scale=0.35]{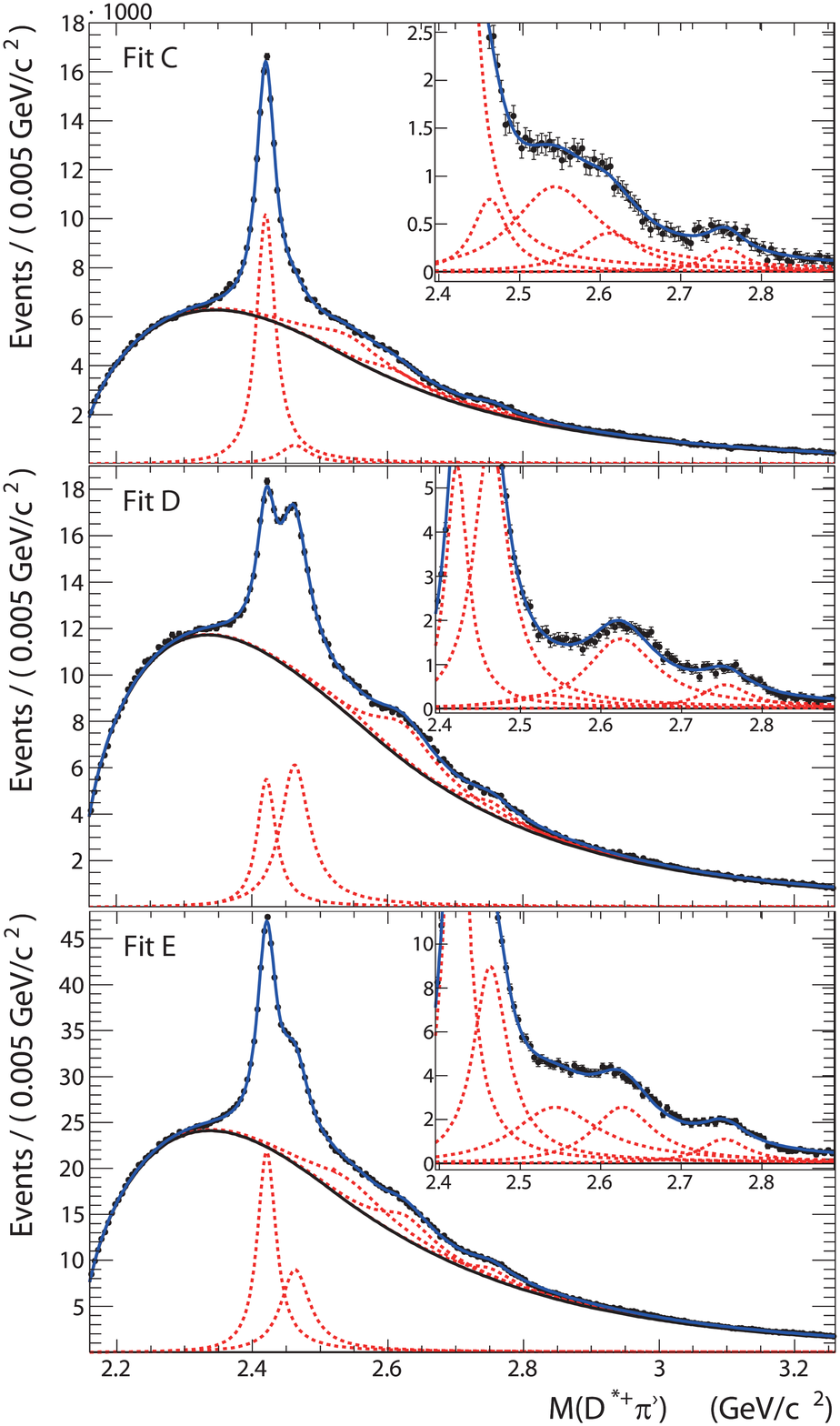}
    \caption{$D^{\ast}(2010)^{+} \pi^{-}$ invariant mass distributions for $|cos \theta_{H}|>0.75$ (top),
             $|cos \theta_{H}|<0.5$ (middle), all candidates (bottom) from~\cite{delAmoSanchez:2010vq}.}
    \label{dstarpi_continuum}
  \end{center}
\end{figure*}

Excited states such as the D-wave or radial excited states are expected in the high mass region.
BaBar studied heavy charmed mesons decaying into $D^{(\ast)} \pi$ produced in the $e^{+} e^{-} \to c \bar{c}$ reaction~\cite{delAmoSanchez:2010vq}.
Figure \ref{dpi_continuum} shows the $D \pi$ invariant mass distribution, where several components are observed
Mesons decaying into $D \pi$ should be a natural parity state. 
The prominent peak around 2460 MeV/$c^{2}$ is due to $D_{2}^{\ast}(2460)$, while the structure around 2300 MeV/$c^{2}$ is 
from the feed-down of $D_{2}^{\ast}(2460)$ or $D_{1}(2420)$ decaying into $D^{\ast} \pi$. 
The two structures around 2.6 GeV/c$^{2}$ and 2.76 GeV/$c^{2}$ originate from $D^{\ast}_{J}(2600)$ and $D^{\ast}_{3}(2750)$,
which are observed for the first time in this analysis. These two states are natural parity states as they decay into $D \pi$.
Figure \ref{dstarpi_continuum} shows the $D^{\ast} \pi$ invariant mass distributions. 
As the spin of $D^{\ast}$ is one, the helicity angle distribution contains spin-parity information of the resonances. 
Here, the helicity angle $\theta_{H}$ is defined as the angle between the primary $\pi$ and the slow $\pi$ in the $D^{\ast}$ rest frame.  
For natural parity states, the decay angle distribution should be $sin^{2} \theta_{H}$
whereas for unnatural parity states, the expected distribution is $cos^{2} \theta_{H}$ for $0^{-}$ 
and $1+hcos^{2} \theta_{H}$ for the other $J^{P}$. Here, $h$ is a helicity parameter.     
The data is divided in two samples:$|cos \theta_{H} > 0.75|$, which enriches the unnatural parity states 
and $|cos \theta_{H} < 0.5|$, which enriches the natural parity states.
In the unnatural parity enriched samples, structures corresponding to $D(2550)$ and $D(2740)$ are visible.
In the natural parity enriched samples, structures corresponding to $D^{\ast}_{J}(2600)$ and $D^{\ast}_{3}(2750)$ are observed.
The decay angular distribution for each resonance is also measured.  $D(2550)$, $D(2740)$, and $D^{\ast}_{J}(2600)$ are consistent 
with $0^{-}$, unnatural parity, and natural parity states, respectively.
These observations suggest that $D^{\ast}_{J}(2600)$ observed in $D \pi$ and $D^{\ast} \pi$ are the same state while $D^{\ast}_{3}(2750)$ and
$D(2740)$ are different states. BaBar also measured ratios of branching fractions: 
\begin{eqnarray}
	{  {\cal B} (D^{\ast}(2600)^{0} \to D^+ \pi^-) \over {\cal B} (D^{\ast}(2600)^{0} \to D^{\ast}(2010)^{+} \pi^-)} &=& (0.32 \pm 0.02 \pm 0.09) \label{new4ratios} \\
	{  {\cal B} (D_{3}^{\ast}(2750)^{0} \to D^+ \pi^- ) \over {\cal B} (D(2740)^{0} \to D^{\ast}(2010)^{+} \pi^-)} &=& (0.42 \pm 0.05 \pm 0.11) \,\,\,\, .
\end{eqnarray}

LHCb experiment studied of $D_{J}^{(\ast)}$ decaying into $D^{(\ast)} \pi$ produced in the $pp$ collision~\cite{Aaij:2013sza}.
The existence of $D(2550)$, $D^{\ast}_{J}(2600)$, $D(2740)$, and $D_{3}^{\ast}(2750)$ are confirmed, but the mass and width of $D^{\ast}_{J}(2600)$
for two measurements are inconsistent. In addition, two states are observed with mass around 3000 MeV/$c^{2}$: $D_{J}(3000)$ and $D_{J}^{\ast}(3000)$,
which decay into $D^{\ast} \pi$ and $D \pi$, respectively. From the helicity angle analysis, $D(2550)$ is consistent with unnatural parity, 
but $0^{-}$ is not preferred over other $J^{P}$.
The resonances in the $D \pi$ final state in $B^{-} \to D^{+} \pi^{-} \pi^{-}$ decay were also studied, and the spin of $D^{\ast}_{3}(2750)$ is determined as three~\cite{Aaij:2015sqa}.

The most straightforward interpretation of $D(2550)$ and $D^{\ast}_{J}(2600)$ is the heavy spin doublet with $n=2$, $L=0$, and $j_{q}=1/2$,
and that for $D(2740)$ and $D^{\ast}(2750)$ is the heavy spin doublet with $n=1$, $L=2$, $j_{q}=5/2$ as $J^{P}$ of $3^{-}$ is
preferred for $D^{\ast}(2750)$. The predicted mass for these quantum numbers are consistent with quark model prediction~\cite{Godfrey:1985xj}.
Many theoretical works studied these states not only from the masses but also from the decay width and the decay branching fractions 
~\cite{Zhong:2010vq,Wang:2010ydc,Colangelo:2012xi,Yu:2014dda}. Calculations with heavy quark effective theory~\cite{Wang:2010ydc} are 
reasonably consistent with these interpretations. The calculation based on the $^{3}P_{0}$ model~\cite{Zhong:2010vq} shows that observed width of $D(2550)$ is too broad
compared to the predicted width of 22.1 MeV. 
Additionally, the LHCb measurement does not prefer $0^{-}$ over other unnatural parity $J^{P}$ for $D(2550)$ (it is called $D(2580)$ in LHCb's measurement~\cite{Aaij:2013sza}).
More efforts theoretically and experimentally are necessary to identify the nature of the observed high mass excited states.

\subsubsection{$D^{0}$ and $D^{\ast 0}$}
BaBar precisely measurement the $D^{\ast}(2010)^{+}$ mass and width by using the decay chain 
$D^{\ast}(2010)^{+} \to D^{0} \pi^{+}$, $D^{0} \to K^{-} \pi^{+}, K^{-} \pi^{+} \pi^{-} \pi^{+}$~\cite{Lees:2013zna}.
The mass and width are obtained as $m(D^{\ast}(2010)^{+}) - m(D^{0}) = 145.4266\pm0.0005\pm0.0020$ MeV/$c^{2}$, and $\Gamma=83.3\pm1.2\pm1.4$ keV,
respectively. This is an improvement of the uncertainty for the width and mass by factors of approximately 12 and 6, respectively.
BaBar precisely measured the $D^{0}$ mass by using the decay chain 
$D^{\ast}(2010)^{+} \to D^{0} \pi^{+}, D^{0} \to K^{-} K^{-} K^{+} \pi^{+}$~\cite{Lees:2013dja}.
The $K^{-} K^{-} K^{+} \pi^{+}$ decay was used as its $Q$-value is a relatively small-value of 250 MeV. 
This yields a small background and a good invariant mass resolution.
The obtained values are $m(D^{0}) = 1864.841\pm0.048\pm0.043 + 3[m(K^{+}) - 493.677]$ MeV/$c^{2}$, where the 
quoted uncertainty is about half of that of previous measurements.
From this measurement, the binding energy of the $X(3872)$ assuming it as $D^{0} \bar{D}^{\ast 0}$ molecular state~\cite{Close:2003sg}
is $0.12\pm0.24$ MeV, indicating $X(3872)$ is a very shallow bound state.

\subsection{Spectroscopy of $D_{s}$ mesons} 
\begin{figure*}[htbp]
    \includegraphics[scale=0.5]{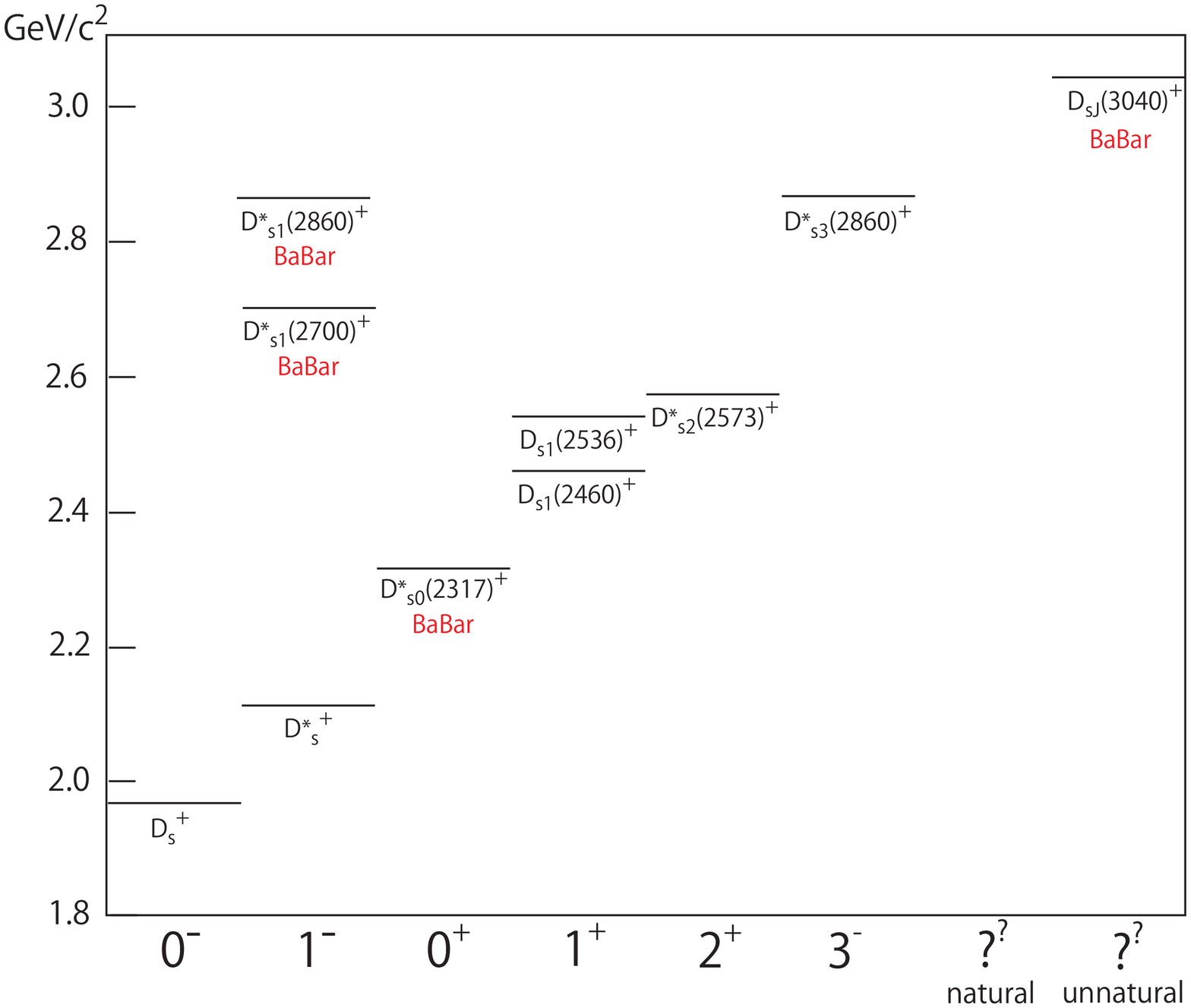}
    \caption{Currently observed $D_{s}$ meson family.  
             For the label of hadrons, convention of Ref.~\cite{Patrignani:2016xqp} is followed.}
    \label{Ds_spectrum}
\end{figure*}

Table \ref{summary_ds} and Fig.~\ref{Ds_spectrum} summarize the observed $D_{s}$ states.
Prior to B-factory experiments, four states were observed.
Two of them were ground state $D_{s}^{+}$ and its heavy quark spin partner $D_{s}^{\ast +}$.
The other two were the heavy quark spin doublet with $L=1$ and $j_{q}=3/2$: $D_{s 1}(2536)^{+}$ and $D_{s 2}^{\ast}(2573)^{+}$.
The B-factory experiments have made significant contributions on $D_{s}$ as five new states have been observed.

\begin{center}
  \begin{table*}[htbp]
     \caption{Summary of the observed $D_{s}^{+}$ family~\cite{Patrignani:2016xqp}. }
     \begin{tabular}{c|cccc} \hline \hline
       Particle                           & Mass (MeV/c$^{2})$     &Width (MeV)          & $J^{P}$     & Strong or EM decay modes                                                                   \\ \hline 
       $D_{s}^{+}$                        & $1968.34\pm0.07$       & -                   & $0^{-}$     &  -                                                                                         \\   
       $D_{s}^{\ast +}$                   & $2112.1\pm0.4$         &$<1.9$               & $1^{-}$     & $D_{S}^{+} \gamma, D_{S}^{+} \pi^{0}$                                                      \\  
       $D_{s 0}^{\ast} (2317)^{+}$        & $2317.7\pm0.6$         &$<3.8$               & $0^{+}$     & $D_{S}^{+} \pi^{0}$                                                                        \\
       $D_{s 1}(2460)^{+} $               & $2459.5\pm0.6$         &$<3.5$               & $1^{+}$     & $D_{S}^{\ast +} \pi^{0}, D_{S}^{+} \gamma, D_{S}^{+} \pi^{+} \pi^{-}$                      \\
       $D_{s 1}(2536)^{+} $               & $2535.10\pm0.06$       &$0.92\pm0.05$        & $1^{+}$     & $D^{\ast +} K^{0}, D^{+} \pi^{-} K^{+}, D^{\ast 0} K^{+}, D^{+}_{S} \pi^{+} \pi^{-}$       \\
       $D_{s 2}^{\ast}(2573)^{+} $        & $2569.1\pm0.8$         &$16.9\pm0.8$         & $2^{+}$     & $D^{0} K^{+}$                                                                              \\
       $D_{s 1}^{\ast}(2700)^{+} $        & $2708.3^{+4.0}_{-3.4}$ &$120\pm11$           & $1^{-}$     & $D^{0} K^{+}, D^{+} K^{0}_{S}, D^{\ast 0} K^{+}, D^{\ast +} K^{0}_{S}$                      \\
       $D_{s 1}^{\ast}(2860)^{+} $        & $2859\pm27$            &$159\pm80$           & $1^{-}$     & $D^{0} K^{+}, D^{+} K_{S}^{0}, D^{\ast 0} K^{+}, D^{\ast +} K^{0}_{S}$                     \\
       $D_{s 3}^{\ast}(2860)^{+} $        & $2860\pm7$             &$53\pm10$            & $3^{-}$     & $D^{0} K^{+} $                                                                             \\
       $D_{s J}(3040)^{+} $               & $3044^{+31}_{-9}$      &$239\pm{60}$         & $?^{?}$     & $D^{\ast 0} K^{+}, D^{\ast +} K^{0}_{S}$                                                   \\ \hline
    \end{tabular}    
    \label{summary_ds}
  \end{table*}
\end{center}

\subsubsection{$D_{s 0}^{\ast} (2317)^{+}$ and $D_{s 1}(2460)^{+} $}
\begin{figure*}[htbp]
  \begin{center}
    \includegraphics[scale=0.55]{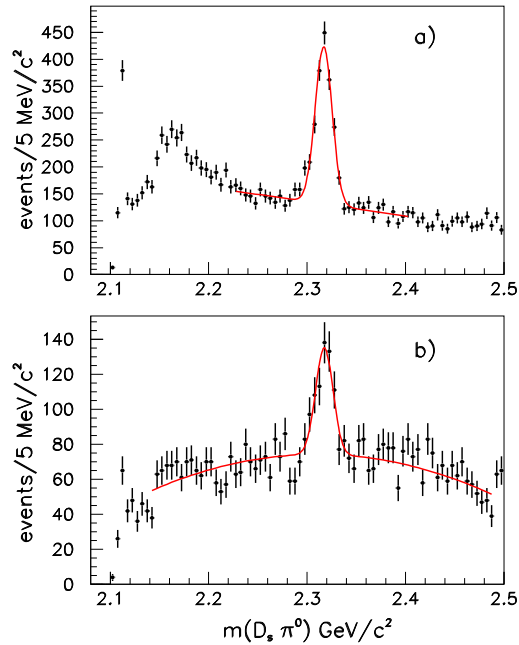}
    \caption{$D_{s}^{+} \pi^{0}$ invariant mass distributions for $D_{s}^{+} \to K^{+} K^{-} \pi^{+}$ (a)
            and $D_{s}^{+} \to K^{+} K^{-} \pi^{+} \pi^{0}$ (b) from~\cite{Aubert:2003fg}.
            Peak corresponding to $D_{s 0}^{\ast} (2317)^{+}$ is observed.}
    \label{ds2317_firstobservation}
  \end{center}
\end{figure*}

For the P-wave excited states, the same discussion as the $D$ family can be applied.
In total, four states are expected. Two have $j_{q}=3/2$ and $J=1,2$. 
They correspond to $D_{s 1}(2536)^{+}$ and $D_{s 2}^{\ast}(2573)^{+}$. The other two states have $j_{q}=1/2$ with 
$J=0,1$. Prior to the B-factory experiments, they had yet to be discovered. The reason for the no-observation were 
thought to be their width were expected to be broad because the expected decay modes were $D K$ ($J=0$) or $D^{\ast} K$ ($J=1$) via the S-wave.  
Therefore, the observation of two very narrow states decaying into $D_{s}^{+} \pi^{0}$ and $D_{s}^{\ast} \pi^{0}$,
which are called $D_{s 0}^{\ast} (2317)^{+}$ and $D_{s 1}(2460)^{+}$, was surprising. 
Figure \ref{ds2317_firstobservation} shows the $D_{s}^{+} \pi^{0}$ invariant mass distributions in the $e^{+} e^{-} \to c\bar{c}$ reaction 
reported by BaBar~\cite{Aubert:2003fg}. This is the first observation of $D_{s 0}^{\ast} (2317)^{+}$.
Additionally, a peaking structure at 2.46 GeV/$c^{2}$ was observed in the $D_{s}^{+} \pi^{0} \gamma$ invariant mass distribution.
Nowadays, the structure is interpreted as $D_{s 1}(2460)^{+}$. However, a detailed study including the determination of the significance was not conducted.
Soon after that, the CLEO experiment confirmed the existence of $D_{s 0}^{\ast} (2317)^{+}$ along with 
the first observation of $D_{s 1}(2460)^{+}$ decaying into $D_{s}^{\ast +} \pi^{0}$~\cite{Besson:2003cp}.
The existence of $D_{s 1}(2460)^{+}$ has been confirmed by Belle~\cite{Abe:2003jk} and BaBar~\cite{Aubert:2003pe}.
Belle also discovered the $D_{s 1}(2460)^{+}$ decay modes of $D_{s}^{+} \gamma$ and $D_{s}^{+} \pi^{+} \pi^{-}$~\cite{Abe:2003jk}.

As the $D_{s 0}^{\ast} (2317)^{+}$ decays into $D_{s}^{+} \pi^{0}$, it is a natural-parity state.
Considering $J^{P}=2^{+}$ state $D_{s 2}^{\ast}(2573)^{+} $ is already observed, the most plausible $J^{P}$ is $0^{+}$.
On the other hand, $D_{s 1}(2460)^{+}$ should be an unnatural-parity state  
because it decays into $D_{s}^{\ast +} \pi^{0}$ and the decay into the isospin allowable mode $D K$ is not observed,
although the mass is above the threshold.
Angular analysis of $D_{s 1}(2460)^{+} \to D_{s}^{+} \gamma$ with $D_{s 1}(2460)^{+}$
produced in the $B$ decay by both Belle and BaBar suggest the spin is one~\cite{Krokovny:2003zq,Aubert:2004pw}.
Therefore, the $J^{P}$ is determined as $1^{+}$.
It is attractive to assign these two states as $j_{q}=1/2$ P-wave states. 
However, the masses of $D_{s 0}^{\ast} (2317)^{+}$ and $D_{s 1} (2460)^{+}$ are smaller than those from
the quark model prediction~\cite{Godfrey:1985xj} by around 160 MeV/$c^{2}$ and 70 MeV/$c^{2}$, respectively.
Actually, the mass of $D_{s 0}^{\ast} (2317)^{+}$ is smaller than its possible non-strange
partner $D_{0}^{\ast}(2400)^{0}$ although the constituent mass of the strange quark is around 100 MeV/$c^{2}$ heavier than up or down quarks.
Additionally, the severe limits for the radiative decay
$\frac{\Gamma(D_{s 0}^{\ast} (2317)^{+} \to D_{s}^{\ast +} \gamma)}{\Gamma(D_{s 0}^{\ast} (2317)^{+} \to D_{s}^{+} \pi^{0})} < 0.059$ by CLEO~\cite{Besson:2003cp}
is inconsistent with the simple $c\bar{s}$ picture~\cite{Godfrey:2003kg}.
These facts have triggered many theoretical discussion for the nature of these two states. 
These include a $D K$ or $D^{\ast} K$ molecule~\cite{Barnes:2003dj}, tetra-quark~\cite{Terasaki:2003qa}, the mixing of $c\bar{s}$ and tetra-quark~\cite{Browder:2003fk}, etc. 
For more details of theoretical discussion, see for example Refs.~\cite{Swanson:2006st,Colangelo:2004vu} and references therein.

In many multiquark scenarios, the isospin is predicted to be one and the existence of doubly charged and neutral partners is suggested, e.g.~\cite{Terasaki:2006qd}.
However, the search for partner states in continuum production by BaBar~\cite{Aubert:2006bk} and in $B$ meson decay by Belle~\cite{Choi:2015lpc}
both found no evidence and the limits on the product of the production cross section and branching fraction are more than one order of magnitude smaller than
those of the charged states. This is a stringent limit for the isospin one scenario interpretation as the cross section for doubly charged and neutral partners 
are naively similar with those of the charged state from the isospin symmetry. 

Both  $D_{s 0}^{\ast} (2317)^{+}$ and $D_{s 1}(2460)^{+}$ have invariant mass distributions consistent with those expected from the detector response
and the upper limit for the width is determined. The obvious reason for the narrow width is that
these decays are isospin-violating modes.
The most severe limits on the width are 3.8 MeV for $D_{s 0}^{\ast} (2317)^{+}$ and 3.5 MeV for $D_{s 1}(2460)^{+}$,
which were determined by BaBar~\cite{Aubert:2006bk}. 
There are many theoretical calculations for the decay width of these states. 
For example, in Ref.~\cite{Colangelo:2003vg}, it is proposed that the isospin-violating decay is originated from 
$\pi^{0}-\eta$ mixing effect and $D_{s 0}^{\ast}$ and the decay width is around 7 keV.  
In Refs.~\cite{Faessler:2007gv,Faessler:2007us}, it is proposed that in $DK$ molecule picture, 
the direct isospin-violating decay is the dominant process as $D$ and $K$ mesons contain (anti) up and down quarks.
The calculated width of $D_{s 0}^{\ast} (2317)^{+}$ is 46.7-111.9 keV depending on the choice of scale parameter.

\begin{figure*}[htbp]
  \begin{center}
    \includegraphics[scale=0.45]{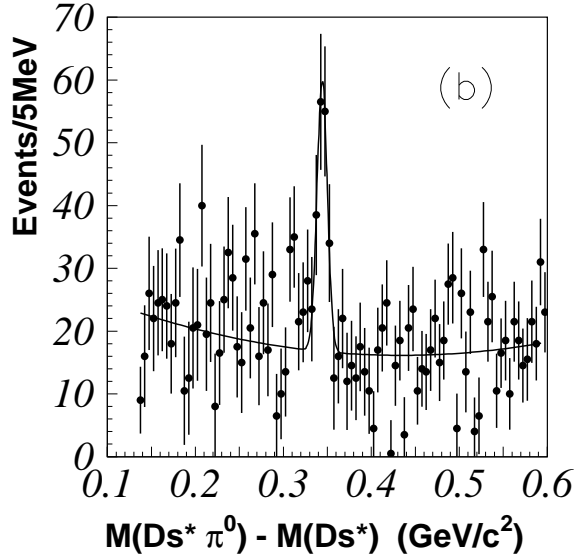}
    \caption{Mass difference of $D_{s}^{\ast +} \pi^{0}$ and $D_{s}^{\ast +}$ showing the observation of 
             $D_{s 1}(2460)^{+}$ by Belle from~\cite{Abe:2003jk}.}
    \label{ds2460_Belle}
  \end{center}
\end{figure*}

\subsubsection{$D_{s 1} (2536)^{+}$}
Observation of the $D_{s 0}^{\ast} (2317)^{+}$ and $D_{s 1}(2460)^{+}$ along with their deviations from the quark model
predictions have renewed interests in the P-wave excited state of the $D_{s}$ family.
The $D_{s 1} (2536)^{+}$, which was first observed by the ARGUS collaboration~\cite{Albrecht:1989yi}, is generally 
accepted as P-wave excited state with  $J=1$ and $j_{q}=1/2$. This is because the mass is very close to the quark model prediction~\cite{Godfrey:1985xj}
and the angular analysis by CLEO~\cite{Alexander:1993nq} as well as the non-observation of the $D K$
decay mode shows it is an unnatural parity state.
In the limit of heavy quark mass to be infinite, states with $j_{q}=1/2$ and $j_{q}=3/2$ must be completely separated. 
However, as the mass of charm quark is finite, $D_{s 1} (2536)^{+}$ should be a mixed state
of the $j_{q}=1/2$ and $j_{q}=3/2$ states. In the  $D_{s 1} (2536)^{+} \to D^{\ast} K$ decay, the 
$j_{q}=1/2$ state decays in the S-wave whereas the $j_{q}=3/2$ state decays in the D-wave. Therefore, 
measuring the relative widths of the D- and S-waves can provide information about the mixing angle. 
Belle performed full angular analysis for the $D_{s 1} (2536)^{+} \to D^{\ast} K, D^{\ast} \to \bar{K} \pi$ decay 
chain with $D_{s 1} (2536)^{+}$ produced in the continuum reaction~\cite{Balagura:2007dya} and showed that 
$\Gamma_{\rm S}/(\Gamma_{\rm S} + \Gamma_{\rm D}) = 0.72\pm0.05\pm0.01$. The S-wave contribution is larger than 
the D-wave one, which is contrary to the naive expectation. 
Before the start of the B-factory experiments, only the upper limit of the width was measured.
BaBar performed the first significant measurement and obtained a value $0.92\pm0.03\pm0.04$ MeV 
using the 8000 $D_{s 1} (2536)^{+}$ sample in the continuum production~\cite{Lees:2011um}.

\subsubsection{Higher excited $D_{s}$ states}
\begin{figure*}[htbp]
  \begin{center}
    \includegraphics[scale=0.3]{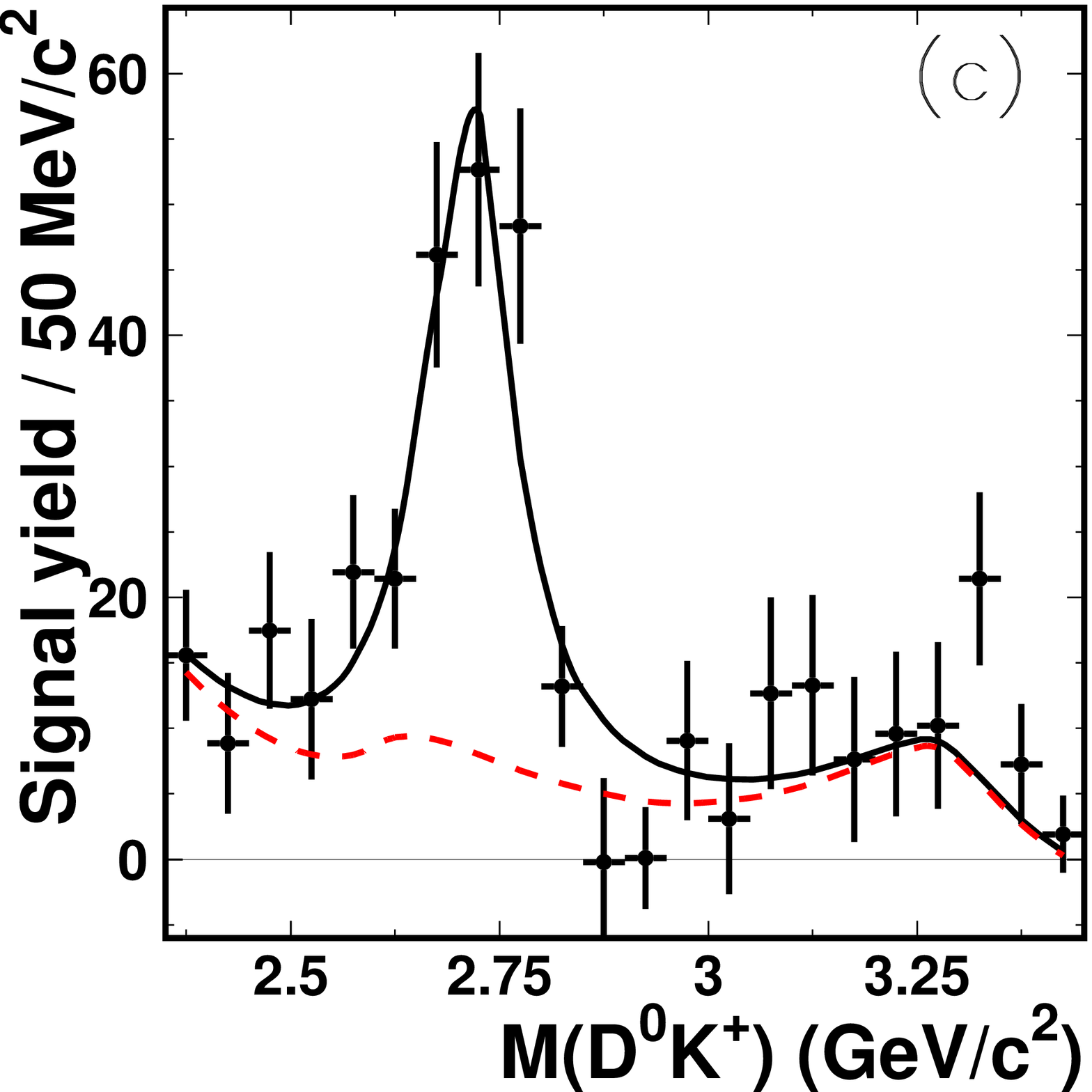}
    \includegraphics[scale=0.3]{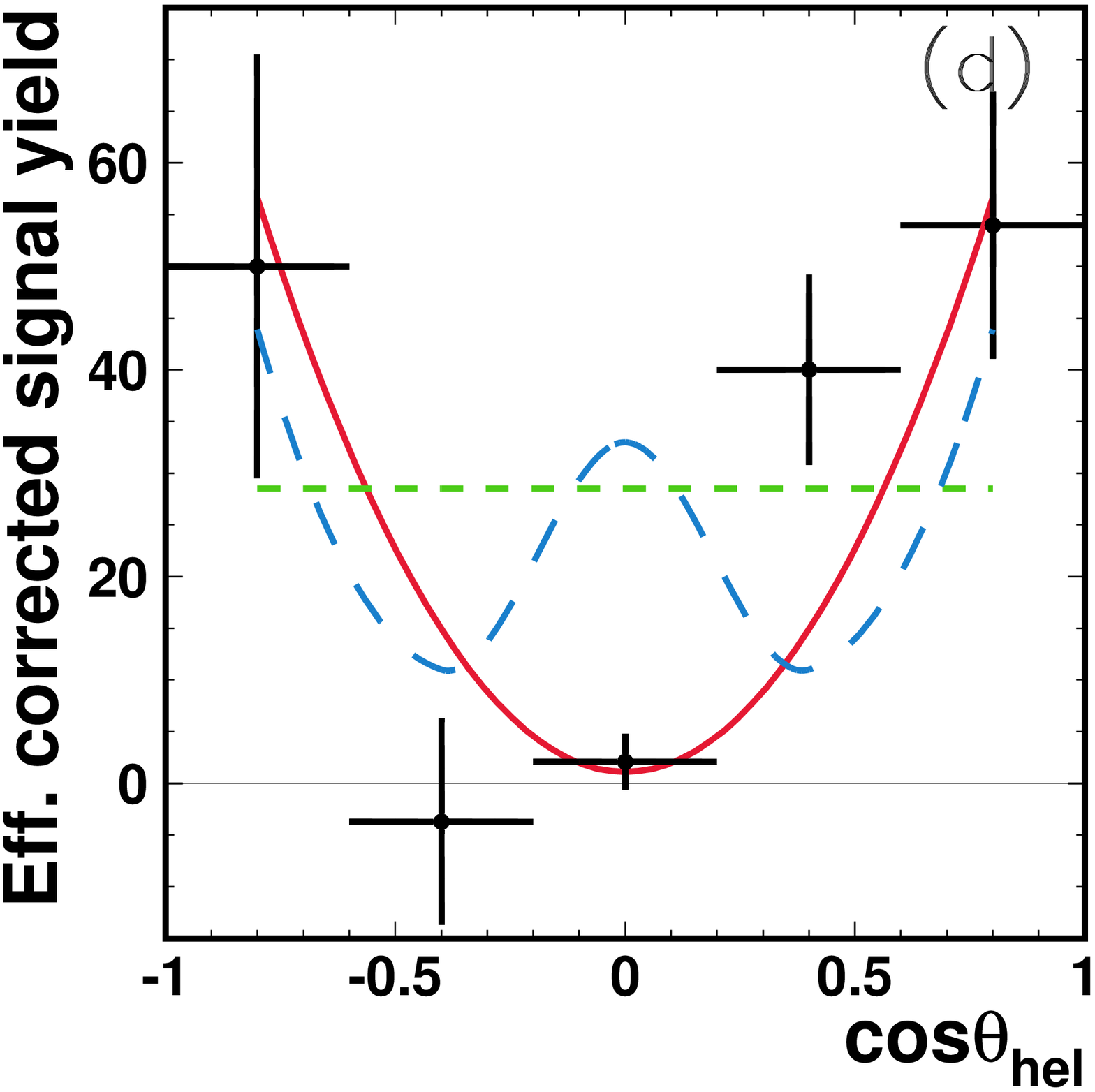}
    \caption{Left: $D^{0} K^{+}$ invariant distribution for $B^{+} \to \bar{D}^{0} D^{0} K^{+}$ decay after removing non-$B$ background. Right: 
	     $D^{\ast}_{s1}(2700)^{+}$ helicity angle distribution. Each curve represents $J=0$ (dotted), $J=1$ (solid), and $J=2$ (dashed) hypothesis, respectively.
             Both plots are from ~\cite{Brodzicka:2007aa}.}
    \label{ds2700_belle}
  \end{center}
\end{figure*}

\begin{figure*}[htbp]
  \begin{center}
    \includegraphics[scale=0.65]{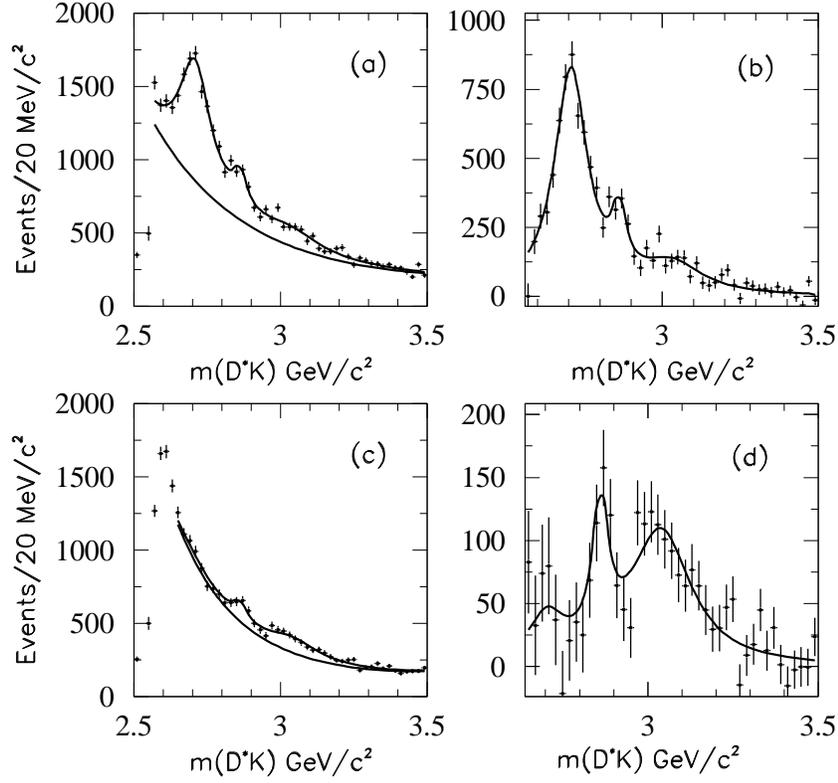}
    \caption{$D^{\ast} K$ invariant mass distributions for (a):$|cos\theta_{H}|<0.4$, (c): $|cos\theta_{H}|>0.4$ from~\cite{Aubert:2009ah}. 
            Here, $\theta_{H}$ is the helicity angle. (b) and (d) show the distribution after the background subtraction.}
    \label{dstark_babar}
  \end{center}
\end{figure*}

BaBar observed a new resonances $D_{s1}^{\ast}(2700)$ and $D_{sJ}^{\ast}(2860)$ decaying into the $D K$ final states 
from continuum production using a data sample of 240 fb$^{-1}$~\cite{Aubert:2006mh}.
The Observation in the $D K$ final states indicates that they are natural parity states.
Soon after, Belle also studied the $D K$ final state in the $B^{+} \to \bar{D}^{0} D^{0} K^{+}$ decay~\cite{Brodzicka:2007aa}.
Belle found a signal with a mass and a width almost consistent with BaBar's result for the $D_{s1}^{\ast}(2700)$ 
with a statistical significance of 8.4$\sigma$. $D_{sJ}^{\ast}(2860)$ observed in the continuum production
is not seen in the $B$ meson decay.
Figure~\ref{ds2700_belle} shows the $D^{0} K^{+}$ invariant mass and the helicity angle distributions. 
From the helicity angle distribution, the spin of the $D_{s1}^{\ast} (2700)^{+}$ is determined as one. 
As the $D K$ decay is possible only for the natural parity state, the $J^{P}$ is determined as $1^{-}$.
BaBar analyzed the $D^{\ast} K$ final state in addition to the $D K$ final state with 
improved statistics of 470 fb$^{-1}$~\cite{Aubert:2009ah}. 
Figure \ref{dstark_babar} shows the $D^{\ast} K$ invariant mass distribution.
In the $D^{\ast} K$ mass distribution, in addition to the $D_{s1}^{\ast} (2700)^{+}$ and $D_{sJ}^{\ast} (2860)^{+}$ observed in the
$D K$ mode, a new resonance $D_{sJ}(3040)^{+}$ is found with a statistical significance of 6.0$\sigma$. 
The non-observation of $D_{sJ}(3040)^{+}$ in the $D K$ decay mode indicates that it is an unnatural parity state.
Additionally, the ratio of branching fractions for $D^{\ast} K$ to $D K$ for 
$D_{s1}^{\ast} (2700)^{+}$ and $D_{sJ}^{\ast} (2860)^{+}$ is reported and the obtained values were almost consistent with one.

LHCb analyzed the $\bar{D}^{0} K^{-}$ final state in the $B_{s} \to \bar{D}^{0} K^{-} \pi^{+}$ decay~\cite{Aaij:2014xza}.
Interestingly, LHCb reported that $D_{sJ}^{\ast}(2860)^{+}$ originated from two overlapped
states with $J^{P}$ equal to $1^{-}$ and $3^{-}$, which we call $D_{s1}^{\ast} (2860)^{+}$ and $D_{s3}^{\ast}(2860)^{+}$.

Both $D_{s1}^{\ast}(2700)^{+}$ and $D_{s1}^{\ast} (2860)^{+}$ have $J^{P}$ equal to $1^{-}$.
In the quark model, $J^{P}=1^{-}$ corresponds to the state with a radially excited state in a S-wave or 
a D-wave excited state with $j_{q}=3/2$. 
In the effective Lagrangian approach, according to the ratio of branching fractions for $D^{\ast} K$ to $D K$,
$D_{s1}^{\ast}(2700)$ is preferred as the radially excited state~\cite{Colangelo:2007ds}. 
It should be noted that as the D-wave and radially excited state should be mixed, 
$D_{s1}^{\ast}(2700)$ and $D_{s1}^{\ast} (2860)^{+}$ should be not a pure radially or D-wave excited state,
but instead are mixed states, The $D_{s3}^{\ast} (2860)^{+}$ is a candidate of the state with a F-wave and $j_{q}=5/2$.
For $D_{sJ}(3040)^{+}$, the experimental information is limited. Hence it is difficult to identify its nature. 


\section{Charmed baryons} 
Charmed baryons are good probes to study the di-quark degree of freedom in the baryons. 
For the wave function of a baryon containing one heavy quark, two Jacobi coordinates can be taken:
the relative coordinate two light quarks ($\rho$) and the coordinates of a di-quark and a charm quark ($\lambda$).
The ratio of the excitation energy for these two coordinates $\frac{\hbar \omega_{\rho}}{\hbar \omega_{\lambda}}$
is given as $\sqrt{\frac{3m_{Q}}{2m_{q}+m{Q}}}$,
where $m_{Q}$ and $m_{q}$ are the constituent masses of the heavy and the light quark, respectively.
In the limit of $m_{Q} \to \infty$, the ratio becomes a large value of $\sqrt{3}$, which means the
excited states for $\rho$ and $\lambda$ modes are separated. A detailed theoretical calculation based on the quark model 
shows the two excitation modes are well separated for the charmed baryon~\cite{Yoshida:2015tia}.
This suggests that charmed baryons can be interpreted rather simply as a bound state of a di-quark and a charm quark.
It is important to verify if this di-quark and charm quark picture systematically explains the experimentally observed spectra. 
Additionally, the discussion of the heavy quark spin symmetry can be also applied to charmed baryons. 
The spectroscopy of charmed baryon provide a information to understand the role of heavy quark in the baryon.

\begin{figure*}[htbp]
    \includegraphics[scale=0.5]{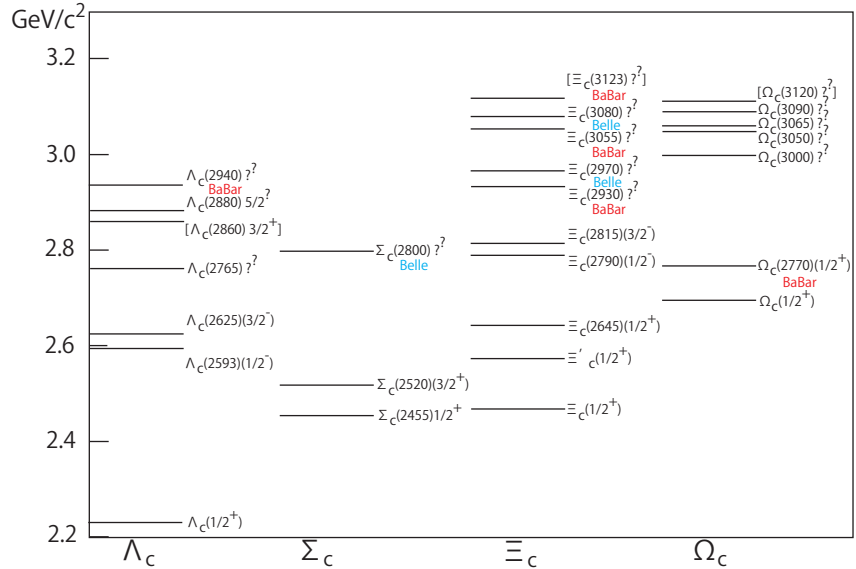}
    \caption{Observed single charmed baryons together with $J^{P}$.   Name of the experiment indicates first observation occurred in Belle or BaBar. 
	     For the label of hadrons, convention of Ref.~\cite{Patrignani:2016xqp} is followed.
             States enclosed in parentheses are observed in a single measurement.
             $J^{P}$ enclosed in parentheses are not from a measurement but from the quark model prediction.}

    \label{charmbaryon_spectrum}
\end{figure*}

Figure \ref{charmbaryon_spectrum} shows the experimentally observed charmed baryons.
Many of the excited states are discovered $e^{+}e^{-}$ collider experiments such as CLEO, Belle, and BaBar.
Historically, CLEO discovered the relatively light excited states while the next generation B-factory experiments 
discovered heavier excited states. Very recently, LHCb also join the spectroscopy of charmed baryons.

\subsection{$\Lambda_{c}^{+}$ family}
$\Lambda_{c}^{+}$ baryons are composed of $cud$ quarks where the isospin of the $ud$ quark system is zero. 
Five $\Lambda_{c}^{+}$ states were observed prior to B-factories. 
The ground state $\Lambda_{c}^{+}$, the heavy quark spin doublet with p-wave ($\lambda$ mode) excited states
$\Lambda_{c}(2593)^{+}$ and $\Lambda_{c}(2625)^{+}$, and the two higher excited states $\Lambda_{c}/\Sigma_{c}(2765)^{+}$ and $\Lambda_{c}(2880)^{+}$. 
For $\Lambda_{c}/\Sigma_{c}(2765)^{+}$, the isospin has yet to be determined.  
Except for the ground state $\Lambda_{c}^{+}$, CLEO discovered these excited states.   

\begin{center}
  \begin{table*}[htbp]
     \caption{Summary of the observed $\Lambda_{c}^{+}$ family~\cite{Patrignani:2016xqp}. Information of $\Lambda_{c}(2860)^{+}$ is from Ref.~\cite{Aaij:2017vbw}. }
     \begin{tabular}{c|cccc} \hline \hline
       Particle                           & Mass (MeV/c$^{2})$      &Width (MeV)    & $J^{P}$   & Observed strong decay modes                        \\ \hline 
       $\Lambda_{c}^{+}$                  & $2286.46\pm0.14$        & -             & $1/2^{+}$ &  -                                                 \\   
       $\Lambda_{c}(2593)^{+}$            & $2592.25\pm0.28$        &$2.6\pm0.6$    & $1/2^{-}$ & $\Lambda_{c}^{+} \pi \pi$, $\Sigma_{c}(2455) \pi$   \\   
       $\Lambda_{c}(2625)^{+}$            & $2628.11\pm0.19$        &$<0.97$        & $3/2^{-}$ & $\Lambda_{c}^{+} \pi \pi$, $\Sigma_{c}(2455) \pi$    \\  
       $\Lambda_{c}/\Sigma_{c}(2765)^{+}$ & $2766.6\pm2.4$          &50             & $?^{?}$   & $\Lambda_{c}^{+} \pi \pi$, $\Sigma_{c}(2455) \pi$, $\Sigma_{c}(2520) \pi$ \\
       $\Lambda_{c}(2860)^{+}$            & $2856.1^{+2.3}_{-6.0}$  &  $68^{+12}_{-22}$ & $3/2^{+}$ & $pD^{0}$  \\
       $\Lambda_{c}(2880)^{+}$            & $2881.63\pm0.24$        &$5.6^{+0.8}_{-0.6}$    & $5/2^{+}$ & $\Lambda_{c}^{+} \pi \pi$, $\Sigma_{c}(2455)$, $\Sigma_{c}(2520) \pi$, $pD^{0}$ \\ 
       $\Lambda_{c}(2940)^{+}$            & $2939.6^{+1.3}_{-1.5}$  &$20^{+6}_{-5}$ & $3/2^{-}$ & $\Sigma_{c}(2455) \pi$, $pD^{0}$ \\ \hline
    \end{tabular}    
    \label{summary_lambdac}
  \end{table*}
\end{center}

\begin{figure*}[htbp]
  \begin{center}
    \includegraphics[scale=0.6]{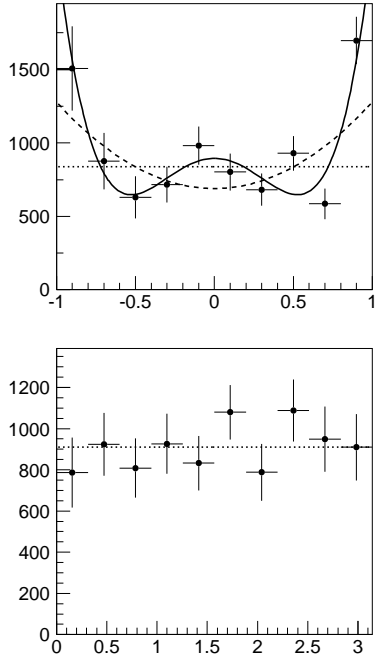}
    \caption{Decay angular distribution of $\Lambda_{c}(2880)^{+} \to \Sigma_{c}(2455) \pi$ as a function of  
            $cos\theta$ and $\phi$ taken from from~\cite{Abe:2006rz}. Solid, dashed, and dotted lines shows the 
            fit results with spin $5/2$, $3/2$, and $1/2$ hypothesis, respectively.}
    \label{lamc2880_decayangle}
  \end{center}
\end{figure*}

Belle studied $J^{P}$ of $\Lambda_{c}(2880)^{+}$ in the $\Sigma_{c}(2455) \pi$ and $\Sigma_{c}(2520) \pi$ final states~\cite{Abe:2006rz}.
Figure \ref{lamc2880_decayangle} shows the decay angular distributions $cos\theta$ and $\phi$, where $\theta$ is the angle
between the pion momentum in the $\Lambda_{c}(2880)^{+}$ rest frame and the boost direction of $\Lambda_{c}(2880)^{+}$ and 
$\phi$ is the angle between the $e^{+} e^{-} \to \Lambda_{c}(2880)^{+} X$ reaction plane and the plane defined by the 
pion momentum and the $\Lambda_{c}(2880)^{+}$ boost direction in the $\Lambda_{c}(2880)^{+}$ rest frame.   
The decay angular distribution for spin $3/2$ hypothesis is given as 
\begin{eqnarray*}
W_{3/2}=\frac{3}{4\pi}[\rho_{33}\sin^2\theta+
\rho_{11}(\frac{1}{3}+\cos^2\theta)-\\
\frac{2}{\sqrt{3}}{\rm Re}\rho_{3-1}\sin^2\theta\cos2\phi-
\frac{2}{\sqrt{3}}{\rm Re}\rho_{31}\sin2\theta\cos\phi]
\end{eqnarray*}
, where $\rho_{ij}$ are the elements of the spin density matrix. The bottom plot of Fig.~\ref{lamc2880_decayangle} shows the 
$\phi$ dependence is small. The decay angular distribution for spin $5/2$  after integrating $\phi$ dependence is given as
\begin{eqnarray*}
W_{5/2}=\frac{3}{8}
[\rho_{55}2(5\cos^4\theta-2\cos^2\theta+1)+\\
\rho_{33}(-15\cos^4\theta+14\cos^2\theta+1)+
\rho_{11}5(1-\cos^2\theta)^2].
\end{eqnarray*}
The decay angular distribution of $\Lambda_{c}(2880)^{+}$ favors the spin $5/2$ hypothesis over the spin $3/2$ and 
$1/2$ at the level of 5.5 and 4.8$\sigma$. The ratio of the decay width for $\Sigma_{c}(2520) \pi$ and $\Sigma_{c}(2455) \pi$ 
is also measured as $R=\Gamma(\Sigma_{c}(2520) \pi)/\Gamma(\Sigma_{c}(2455) \pi)=0.225\pm0.062\pm0.025$.
This value favors $J^{P}=5/2^{+}$ over $J^{P}=5/2^{-}$ from the calculations assuming heavy quark spin symmetry~\cite{Cheng:2006dk}.
There is a subtlety in this calculation as the $P$-wave component in the $\Sigma_{c}(2520) \pi$ decay mode is neglected 
for $5/2^{+}$ assumption. A calculation based on the quark model incorporating heavy quark spin symmetry~\cite{Nagahiro:2016nsx}
shows that in a certain configuration, the $P$-wave contribution can be neglected and $J^{P}=5/2^{+}$ gives $R$ consistent with
the measured value. 

\begin{figure*}[htbp]
  \begin{center}
    \includegraphics[scale=0.5]{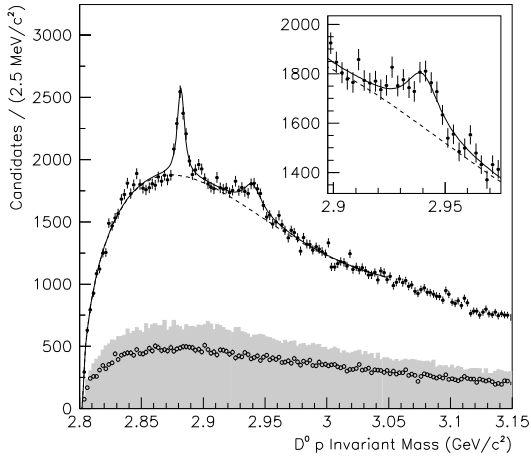}
    \caption{$D^{0} p$ invariant mass distribution from \cite{Aubert:2006sp}.
             Clear peaks corresponding to $\Lambda_{c}(2880)^{+}$ and $\Lambda_{c}(2940)^{+}$ are observed.}
    \label{md0p}
  \end{center}
\end{figure*}

BaBar discovered $\Lambda_{c}(2940)^{+}$ in the $D^{0} p$ final state \cite{Aubert:2006sp}.
Figure \ref{md0p} shows the $D^{0} p$ invariant mass distribution. Clear peaks corresponding to 
$\Lambda_{c}(2940)^{+}$ is observed in addition to well known $\Lambda_{c}(2880)^{+}$ with a significance of 8.7$\sigma$.
As the doubly charged partner for $\Lambda_{c}(2880)^{+}$ and $\Lambda_{c}(2940)^{+}$ is not observed in the $D^{+} p$ final state,
this state is identified as an isosinglet.  These are the firstly observed charmed baryons 
in the decay mode where the charm quark is contained in the meson ($D^{0}$). 

LHCb studied the decay $\Lambda_{b}^{0} \to D^{0} p \pi^{-}$ and reported the existence of $\Lambda_{c} (2860)^{+}$ 
in addition to known $\Lambda_c(2880)^{+}$ and $\Lambda_{c}(2940)^{+}$~\cite{Aaij:2017vbw}. Constraining on $J^{P}$ of these states,
$J^{P}$ of $\Lambda_{c}(2880)^{+}$ is $J=5/2$, with the $J=7/2$ disfavored by 4$\sigma$ and $1/2$ and $3/2$ disfavored 
by more than 5$\sigma$, which is consistent with Belle's result~\cite{Abe:2006rz}. 
$J^{P}$ of $\Lambda_{c}(2860)^{+}$ is identified as $3/2^{+}$ by excluding other quantum numbers with more than 6$\sigma$.
$\Lambda_{c}(2940)^{+}$ favors $3/2^{-}$ but other solutions with spin $1/2$ and $7/2$ cannot be excluded. 

From the facts that the masses of $\Lambda_{c}(2860)^{+}$ and $\Lambda_{c}(2880)^{+}$ are close,
and the measured $J^{P}$ of $3/2^{+}$ and $5/2^{+}$,
it is natural to regard these states as a heavy quark spin doublet with D-wave excitation.
Theoretical analysis with the QCD sum rule suggests that D-wave excitation in the $\lambda$ mode is
consistent with the measured masses~\cite{Wang:2017vtv}.
These predictions for the masses are consistent with several previous theoretical studies~\cite{Ebert:2011kk,Ebert:2011kk,Shah:2016mig} .
However, the authors in Ref.~\cite{Chen:2017aqm} say there are
some difficulties to interpret $\Lambda_{c}(2880)^{+}$ as D-wave excited state from their decay information.

As the mass of $\Lambda_{c}(2940)^{+}$ is very close to the $D^{\ast} p$ threshold (2945 MeV/$c^{2}$), there are a number of 
theoretical papers discussing the possible molecular state interpretation~\cite{He:2006is,Dong:2011ys,Zhang:2012jk,Ortega:2012cx,He:2010zq}.
There are also a number of calculations trying to interpret $\Lambda_{c}(2940)^{+}$ as an ordinal charmed baryon~\cite{Cheng:2006dk,Chen:2007xf}.
However, its nature has yet to be conclusively identified.


\begin{figure*}[htbp]
  \begin{center}
    \includegraphics[scale=0.4]{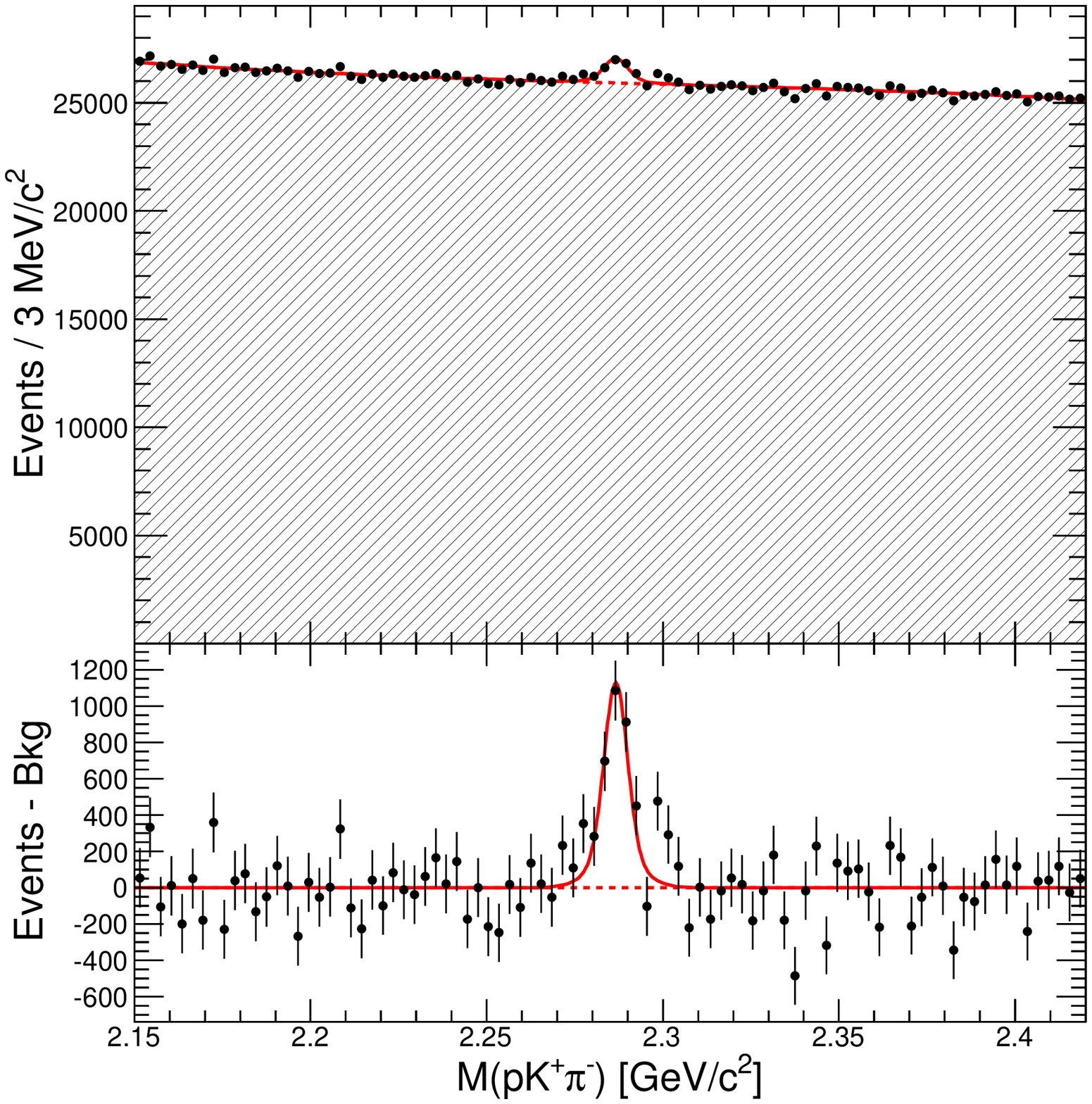}
    \caption{$p K^{+} \pi^{-}$ invariant mass distribution (top) and the same after the background subtraction (bottom) taken from \cite{Yang:2015ytm}.}
    \label{dcs}
  \end{center}
\end{figure*}

The properties of the ground state $\Lambda_{c}^{+}$ are also extensively studied by the B-factory experiments.  
BaBar precisely measured the mass of $\Lambda_{c}^{+}$ using two decay modes $\Lambda K^{0}_{S} K^{+}$
and $\Sigma^{0} K^{0}_{S} K^{+}$, which have very small $Q$ value~\cite{Aubert:2005gt}. 
The result is $m(\Lambda_{c}^{+}) = 2286.46 \pm 0.14$ MeV/c$^{2}$,
which is better than previous measurements by more than factor four, and is still the best measurement. 
Belle performed a model independent measurement of ${\cal B}(\Lambda_{c}^{+} \to pK^{-} \pi^{+})$
by inclusively reconstructing $\Lambda_{c}^{+}$ by using the missing mass recoiling against $D^{\ast-} \bar{p} \pi^{+}$ system~\cite{Zupanc:2013iki}.
The obtained branching fraction $6.84\pm0.24 (stat.)^{+0.21}_{-0.27} (syst.)$ has five times better accuracy than previous measurements. 
Belle also observed the first doubly Cabibbo Suppressed decay of baryon $\Lambda_{c}^{+} \to p K^{+} \pi^{-}$~\cite{Yang:2015ytm}. 
Figure \ref{dcs} shows the $p K^{+} \pi^{-}$ invariant mass distribution. The clear peak corresponds to $\Lambda_{c}^{+}$.
The ratio of the branching fractions $\frac{{\cal B}(\Lambda_{c}^{+} \to pK^{+} \pi^{-})}   { {\cal B}(\Lambda_{c}^{+} \to pK^{-} \pi^{+}) }$ 
is obtained as $(2.35\pm0.27\pm0.21)\times 10^{-3}$. 

\subsection{$\Sigma_{c}$ family}
$\Sigma_{c}$ baryons are composed of the $c$ quark and two $ud$ quarks with an isospin one. 
Table~\ref{summary_sigmac} summarizes the experimentally observed $\Sigma_{c}$ family.
The ground state has $j_{q}=1$ in order to make their wave function totally anti-symmetric. Consequently, a heavy quark spin doublet of $\Sigma_{c}(2455)$ and $\Sigma_{c}(2520)$ is formed. They were the only observed states prior to B-factory experiments.
\begin{center}
  \begin{table*}[htbp]
     \caption{Summary of the observed $\Sigma_{c}^{+}$ family~\cite{Patrignani:2016xqp}. 
              For $\Sigma_{c}(2800)^{0}$, peaks observed in Ref.~\cite{Mizuk:2004yu} and Ref.~\cite{Aubert:2008ax} are assumed to be the same.}
     \begin{tabular}{c|cccc} \hline \hline
       Particle                           & Mass (MeV/c$^{2}$)        &Width (MeV)                 & $J^{P}$   & Observed strong decay modes     \\ \hline 
       $\Sigma_{c}(2455)^{++}$            & $2453.97\pm0.14$          & $1.89^{+0.09}_{-0.18}$     & $1/2^{+}$ & $\Lambda_{c}^{+} \pi^{+}$      \\   
       $\Sigma_{c}(2455)^{+}$             & $2452.9\pm0.4$            & $<4.6$                     & $1/2^{+}$ & $\Lambda_{c}^{+} \pi^{0}$      \\   
       $\Sigma_{c}(2455)^{0}$             & $2453.75\pm0.14$          & $1.83^{+0.11}_{-0.19}$     & $1/2^{+}$ & $\Lambda_{c}^{+} \pi^{-}$      \\   \hline

       $\Sigma_{c}(2520)^{++}$            & $2518.41^{+0.21}_{-0.19}$ & $14.78^{+0.30}_{-0.40}$    & $3/2^{+}$ & $\Lambda_{c}^{+} \pi^{+}$      \\   
       $\Sigma_{c}(2520)^{+}$             & $2517.5\pm2.3$            & $<17$                      & $3/2^{+}$ & $\Lambda_{c}^{+} \pi^{0}$      \\   
       $\Sigma_{c}(2520)^{0}$             & $2518.48\pm0.20$          & $15.3^{+0.4}_{-0.5}$       & $3/2^{+}$ & $\Lambda_{c}^{+} \pi^{-}$      \\  \hline 

       $\Sigma_{c}(2800)^{++}$            & $2801^{+4}_{-6}$          & $75^{+22}_{-17}$     & $?^{?}$ & $\Lambda_{c}^{+} \pi^{+}$        \\   
       $\Sigma_{c}(2800)^{+}$             & $2792^{+14}_{-5}$         & $62^{+60}_{-40}$     & $?^{?}$ & $\Lambda_{c}^{+} \pi^{0}$        \\   
       $\Sigma_{c}(2800)^{0}$             & $2806^{+5}_{-7}$          & $72^{+22}_{-15}$     & $?^{?}$ & $\Lambda_{c}^{+} \pi^{-}$        \\  \hline
    \end{tabular}    
    \label{summary_sigmac}
  \end{table*}
\end{center}

\begin{figure*}[htbp]
  \begin{center}
    \includegraphics[scale=0.6]{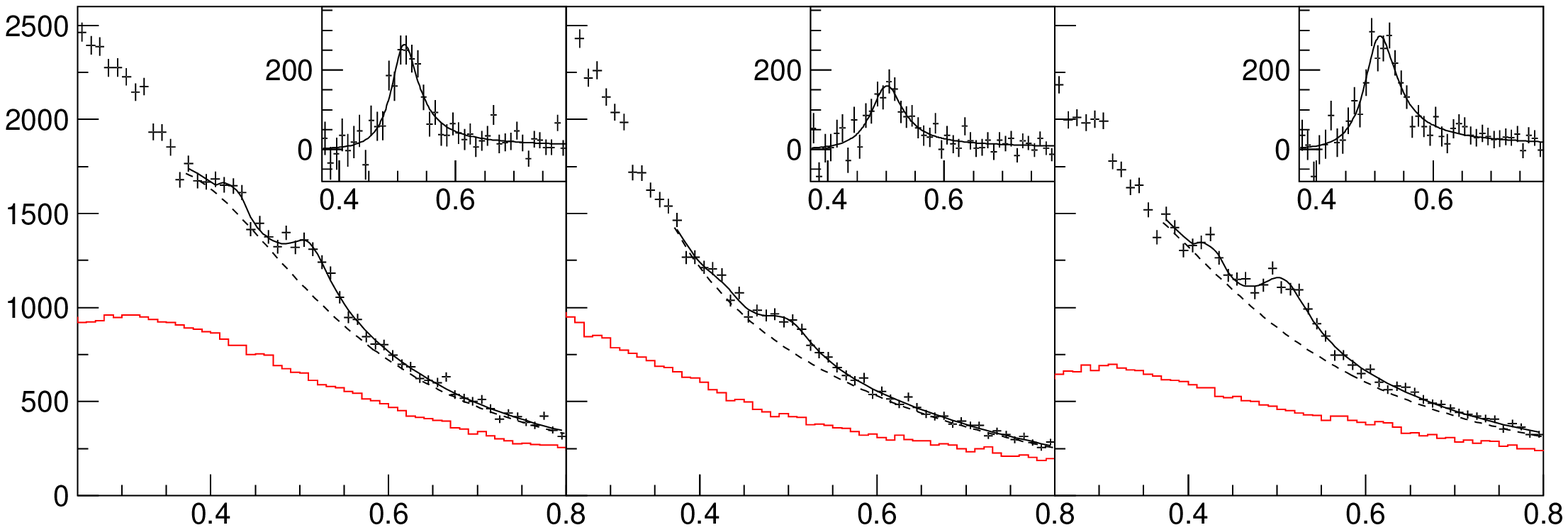}
    \caption{$M(\Lambda_{c}^{+} \pi) - M(\Lambda_{c}^{+})$ distributions with $\Lambda_{c}^{+} \pi^{+}$ (left)
             $\Lambda_{c}^{+} \pi^{0}$ (middle) $\Lambda_{c}^{+} \pi^{-}$ (right) from from~\cite{Mizuk:2004yu}. 
             Peaks corresponding to $\Sigma_{c}(2800)$ are observed.}
    \label{sigc2800}
  \end{center}
\end{figure*}

Belle observed new excited state $\Sigma_{c}(2800)$ in the $\Lambda_{c}^{+} \pi$ decay~\cite{Mizuk:2004yu}.
Figure \ref{sigc2800} shows the $M(\Lambda_{c}^{+} \pi) - M(\Lambda_{c}^{+})$ distributions for three pion charges.
All the isotriplet states are observed with the masses and widths close to each other.
It is noteworthy that we do not see a signal corresponding to $\Lambda_{c}/\Sigma_{c}(2765)$ in this channel,   
suggesting that this state is an isosinglet. $J^{P}$ of $\Sigma_{c}(2800)$ was not measured, making it difficult to identify the nature.
One of the candidates may be the P-wave excitation in the $\lambda$ mode as it is the lightest excited state.
Refs.~\cite{Cheng:2006dk,Chen:2007xf} identified it as $J^{P}=3/2^{-}$ or $5/2^{-}$ in the $\lambda$ mode P-wave excitation
because it decay into $\Lambda_{c}^{+} \pi$ with D-wave and the width can be relatively narrow.
There are many other studies such as coupled channel~\cite{Mizutani:2006vq}, and the $DN$ molecule state~\cite{Dong:2010gu,Zhang:2012jk} ($DN$ threshold is 2804 MeV/$c^{2}$),

BaBar studied the intermediate $\Lambda_{c}^{+} \pi^{-}$ final state in the $B^{-} \to \Lambda_{c}^{+} \bar{p} \pi^{-}$ decay~\cite{Aubert:2008ax}.
An enhancement with mass of $2846\pm8$ MeV/c$^{2}$ is observed. This value is 3$\sigma$ higher than that of 
$\Sigma_{c}(2800)^{0}$ measured by Belle~\cite{Mizuk:2004yu} (widths in the two measurements are consistent each other).
A study with higher statistics is needed to resolve whether these two peaks are originated from the same excited state.
BaBar also determined the spin of the $\Sigma_{c}(2455)^{0}$ from the decay angular distribution in the $B^{-} \to \Sigma_{c}(2455)^{0} \bar{p}$ decay.
The spin $1/2$ is preferred over $3/2$ at the 4$\sigma$ level, which is consistent with the quark model expectation for 
the lowest $\Sigma_{c}$ state. In this decay mode, the helicity of $\Sigma_{c}(2455)$ is fixed to be $1/2$ as the 
angular spin of the $B$ meson is zero and that for a proton is $1/2$. This study clearly shows the usefulness 
of the $B$ meson decay to determine the quantum number of the charmed baryons.
It is important to measure the mass and width with a good precision to conclude if the two states observed at Belle and BaBar are the same.

The mass difference of the iso-multiplet is interesting as it contains information on the wave function of the charmed baryons.
Naively, the mass of the neutral state is heavier than the doubly charged one because $m(d)$ is heavier than $m(u)$. 
However, due to the electro-magnetic repulsive force, the doubly charged state is predicted to be heavier than neutral one~\cite{Chan:1985ty,Hwang:1986ee,Capstick:1987cw,SilvestreBrac:2003kd}.
Belle precisely measured for the masses and widths of $\Sigma_{c}(2455/2520)$ using full statistics~\cite{Lee:2014htd}.
The obtained values are $m(\Sigma_{c}(2455)^{++}) - m(\Sigma_{c}(2455)^{0}) = 0.22\pm0.01\pm0.01$ MeV/$c^{2}$ and 
$m(\Sigma_{c}(2520)^{++}) - m(\Sigma_{c}(2520)^{0}) = 0.01\pm0.15\pm0.03$ MeV/$c^{2}$. 
For $\Sigma_{c}(2455)$, the obtained mass of the doubly charged state is significantly heavier than that of the neutral state.

\subsection{$\Xi_{c}$ family}
\begin{figure*}[htbp]
  \begin{center}
    \includegraphics[scale=0.3]{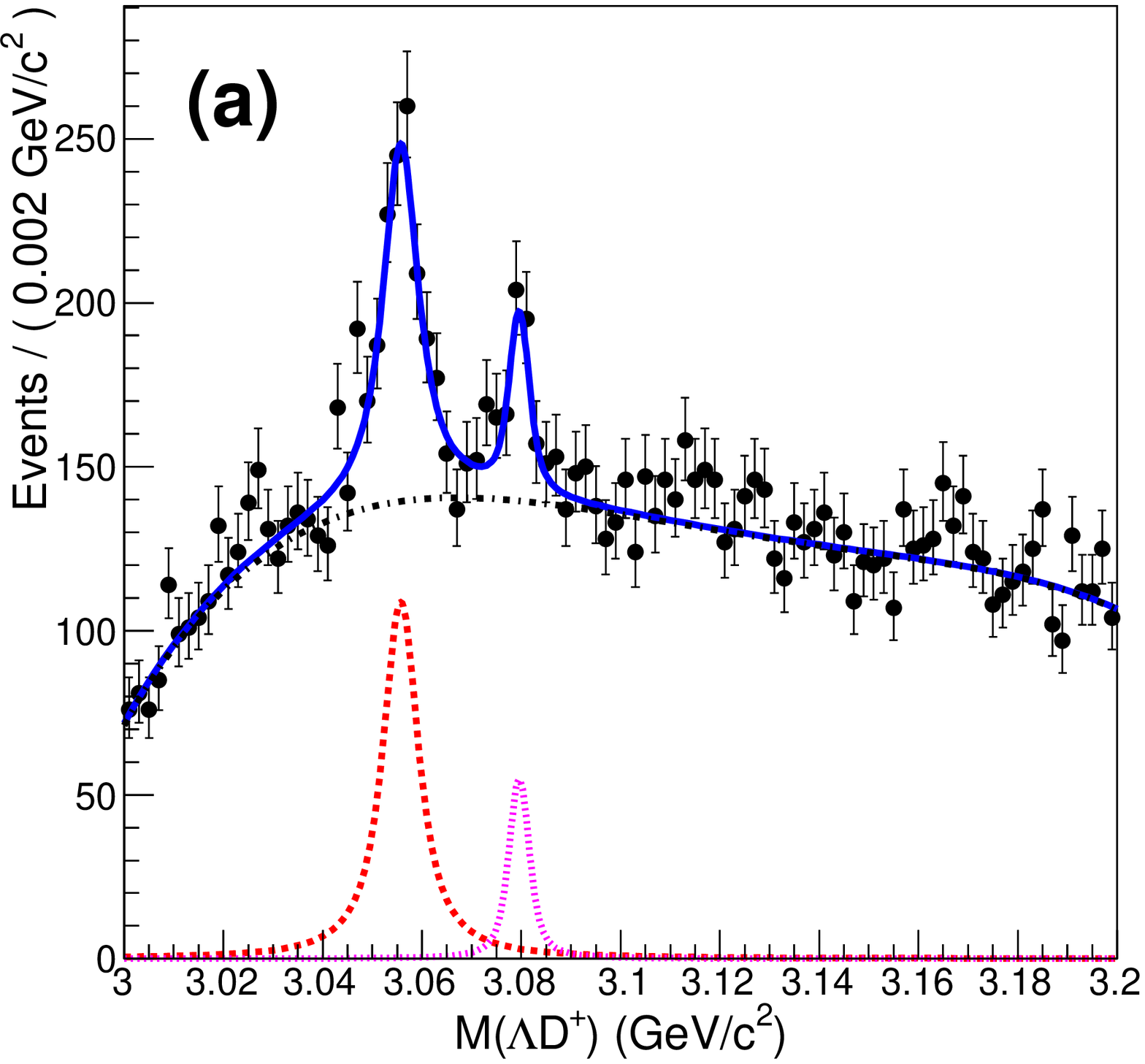}
    \includegraphics[scale=0.3]{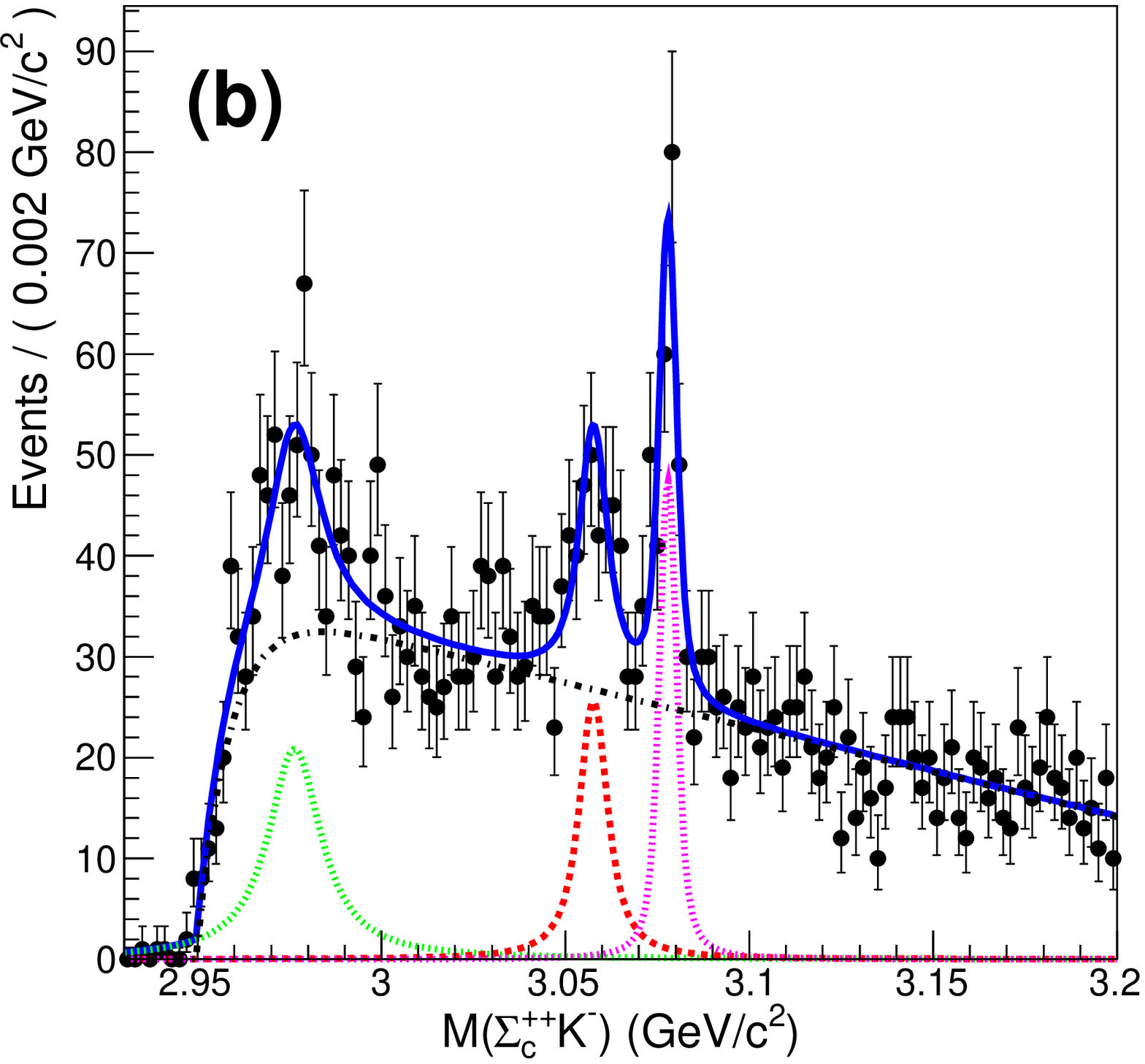}
    \includegraphics[scale=0.3]{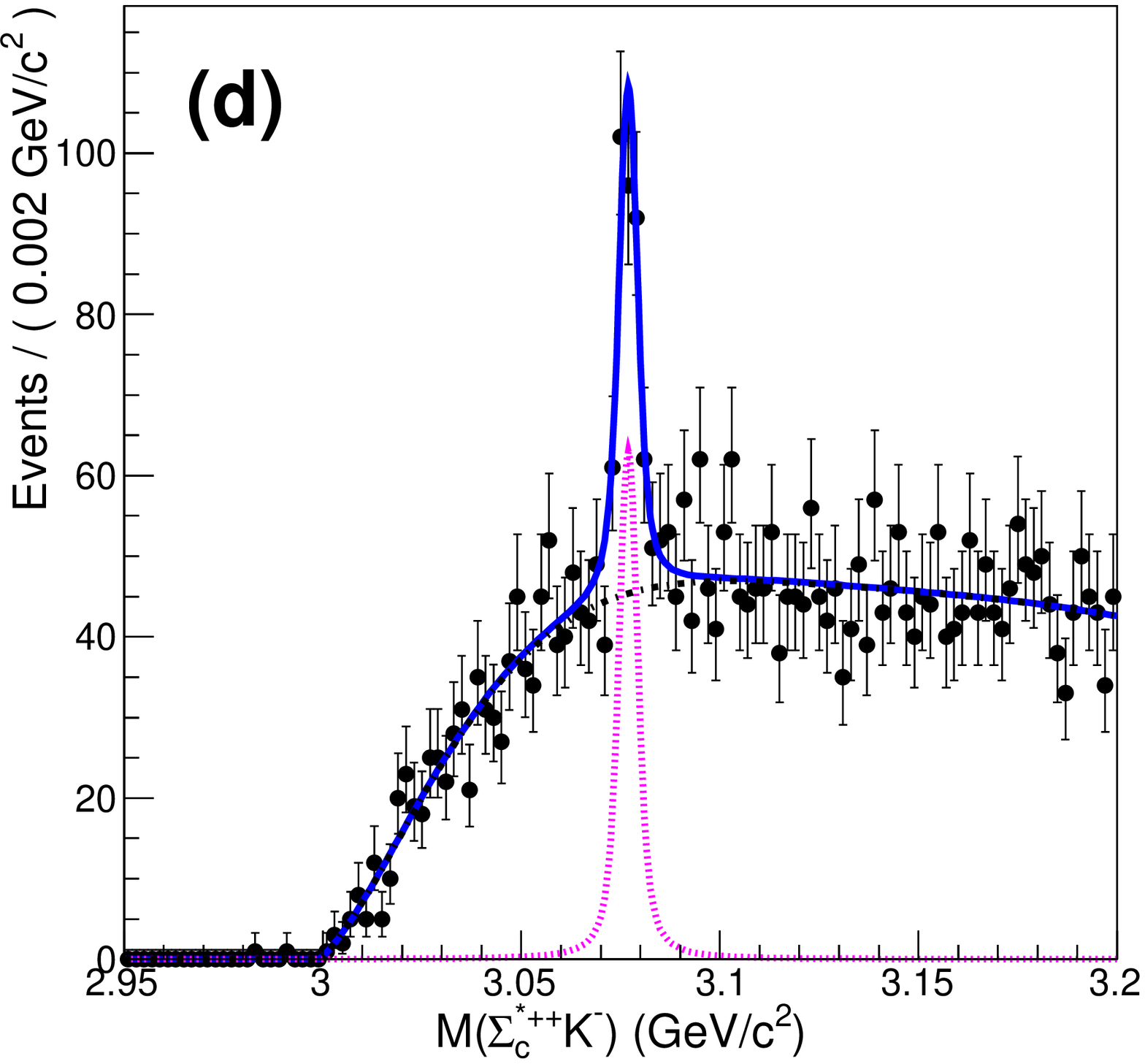}
    \caption{Invariant mass distributions of $\Lambda D^{+}$ (left), $\Sigma_{c}(2455)^{++} K^{-}$ (middle), and $\Sigma_{c}(2520)^{++}K^{-}$ (right),
             taken from~\cite{Kato:2016hca}.}
    \label{xics}
  \end{center}
\end{figure*}

$\Xi_{c}$ baryons are composed of $cs$ quarks and one $u$ or $d$ quark. 
Table~\ref{summary_guzaic} summarizes the experimentally observed $\Xi_{c}$ family.
Five $\Xi_{c}$ states were observed prior to the B-factory experiments:ground state $\Xi_{c}$, heavy quark spin doublet with $su/d$ di-quark to be one,
$\Xi_{c}^{'}$, $\Xi_{c}(2645)^{+}$, another heavy quark spin doublet with a P-wave in the $\lambda$ mode excitation,
$\Xi_{c}(2790)$, and $\Xi_{c}(2815)$. 
The B-factory experiments have established three new excited $\Xi_{c}$ states.
\begin{center}
  \begin{table*}[htbp]
	  \caption{Summary of the observed $\Xi_{c}$ family~\cite{Patrignani:2016xqp} Information of $\Xi_{c}(3055)^{0}$ is from ~\cite{Kato:2016hca}.
                   States enclosed with parentheses shows those observed with only one measurement.}
     \begin{tabular}{c|cccc}     \hline \hline
       Particle              & Mass (MeV/c$^{2}$)        &Width (MeV)           & $J^{P}$      & Strong/EM decay modes                        \\ \hline 
       $\Xi_{c}^{+}$         & $2467.87\pm0.3$           & -                    & $1/2^{+}$    & -                                            \\   
       $\Xi_{c}^{0}$         & $2470.87^{+0.28}_{-031}$  & -                    & $1/2^{+}$    & -                                            \\ \hline

       $\Xi_{c}^{'+}$        & $2577.4 \pm 1.2$          & -                    & $1/2^{+}$    & $\Xi_{c}^{'+} \gamma$                        \\   
       $\Xi_{c}^{'0}$        & $2578.8 \pm 0.5$          & -                    & $1/2^{+}$    & $\Xi_{c}^{'0} \gamma$                        \\ \hline 

       $\Xi_{c}(2645)^{+}$   & $2645.53 \pm 0.31$        & $2.14\pm0.19$        & $3/2^{+}$    & $\Xi_{c}^{0} \pi^{+}$                        \\   
       $\Xi_{c}(2645)^{0}$   & $2646.32 \pm 0.31$        & $2.35\pm0.22$        & $3/2^{+}$    & $\Xi_{c}^{+} \pi^{-}$                        \\ \hline  

       $\Xi_{c}(2790)^{+}$   & $2792.0 \pm 0.5$          & $8.9\pm1.0$          & $1/2^{-}$    & $\Xi^{'}_{c}  \pi^{+}$                       \\   
       $\Xi_{c}(2790)^{0}$   & $2792.8 \pm 1.2$          & $10.0\pm1.1$         & $1/2^{-}$    & $\Xi_{c}^{'+} \pi^{-}$                        \\ \hline  

       $\Xi_{c}(2815)^{+}$   & $2816.67 \pm 0.31$        & $2.43\pm0.26$        & $3/2^{-}$    & $\Xi_{c} \pi, \Xi_{c}^{+} \pi^{+} \pi^{-}, \Xi_{c}(2645)^{+} \pi^{-}, \Xi^{'}_{c}  \pi^{+}$   \\   
       $\Xi_{c}(2815)^{0}$   & $2820.22 \pm 0.32$        & $2.54\pm0.25$        & $3/2^{-}$    & $\Xi_{c} \pi, \Xi_{c}^{0} \pi^{+} \pi^{-}, \Xi_{c}(2645)^{0} \pi^{+}  \Xi_{c}^{'+} \pi^{-}$   \\ \hline  

       $\Xi_{c}(2930)^{0}$   & $ 2931\pm 6$              & $36\pm13$            & $?^{?}$      & $\Lambda_{c}^{+} \bar{K}$     \\ \hline  

       $\Xi_{c}(2970)^{+}$   & $2969.5 \pm 0.8$          & $20.9^{+2.4}_{-3.5}$ & $?^{?}$      & $\Lambda_{c}^{+} K^{-} \pi^{+}, \Sigma_{c}(2455)^{++}K^{-}, \Xi_{c}^{+} \pi^{+} \pi^{-}, \Xi_{c}(2645)^{0}\pi^{+}$ \\  
       $\Xi_{c}(2970)^{0}$   & $2967.8 \pm 0.8$          & $28.1^{+3.4}_{-4.0}$ & $?^{?}$      & $\Lambda_{c}^{+} \bar{K}^{0} \pi^{+}, \Xi_{c}(2645)^{+} \pi^{-}$   \\  \hline 

       $\Xi_{c}(3055)^{+}$   & $3055.9 \pm 0.4$          & $7.8\pm1.9$    & $?^{?}$            & $\Sigma_{c}(2455)^{++} K^{-}, \Lambda D^{+}$   \\   
       ($\Xi_{c}(3055)^{0}$) & $3059.0 \pm 0.5\pm0.6$    & $6.4\pm2.1\pm1.1$    & $?^{?}$      & $\Lambda D^{0}$   \\  \hline 

       $\Xi_{c}(3080)^{+}$   & $3077.2 \pm 0.4$          & $3.6\pm1.1$          & $?^{?}$      & $\Lambda_{c}^{+} K^{-} \pi^{+}, \Sigma_{c}(2455)^{++}K^{-}, \Sigma_{c}(2520)^{++} K^{-}, \Lambda D^{+}$   \\   
       $\Xi_{c}(3080)^{0}$   & $3079.9 \pm 1.4$          & $5.6\pm2.2$          & $?^{?}$      & $\Lambda_{c}^{+} \bar{K}^{0} \pi^{+}, \Sigma_{c}(2455)^{0} \bar{K}^{0}, \Sigma_{c}(2520)^{0} \bar{K}^{0}$ \\   \hline

       ($\Xi_{c}(3123)^{+}$) & $3122.9 \pm 1.3\pm0.3$    & $4.4\pm3.4\pm1.7$    & $?^{?}$   & $ \Sigma_{c}(2520)^{++} K^{-}$                \\ \hline

    \end{tabular}    
    \label{summary_guzaic}
  \end{table*}
\end{center}
Three papers have reported the excited $\Xi_{c}$ states decaying into $\Lambda_{c}^{+} K^{-} \pi^{+}$ final states from B-factories.
Belle reported observation of the excited $\Xi_{c}$ states of $\Xi_{c}(2970)^{+}$ and $\Xi_{c}(3080)^{+/0}$ in the 
$\Lambda_{c}^{+} K^{-} \pi^{+}$ and $\Lambda_{c}^{+} K_{S}^{0} \pi^{-}$ final states~\cite{Chistov:2006zj} with data sample of 461.5 fb$^{-1}$.
BaBar studied the same final states, as well as  $\Sigma_{c}(2455)$ and $\Sigma_{c}(2520)$ into $\Lambda_{c}^{+} \pi$
intermediate states~\cite{Aubert:2007dt}.
In addition to the $\Xi_{c}(2970)$ and $\Xi_{c}(3080)$, BaBar discovered $\Xi_{c}(3055)^{+}$ in the $\Sigma_{c}(2455)^{++} K^{-}$
final state and found a evidence of $\Xi_{c}(3123)^{+}$ in the $\Sigma_{c}(2520)^{++} K^{-}$ final state.
Finally, Belle studied the $\Sigma_{c}(2455)^{++} K^{-}$ and $\Sigma_{c}(2520)^{++} K^{-}$
final state with a data sample of 980 fb$^{-1}$~\cite{Kato:2013ynr}. The existence of $\Xi_{c}(3055)^{+}$ is confirmed, 
but a peak structure corresponding to $\Xi_{c}(3123)^{+}$ is not observed.
The $\Xi_{c}(2970)^{+/0}$ is also observed in the $\Xi_{c}(2645)^{0/+} \pi^{+/-}$ final state~\cite{Lesiak:2008wz} and $\Xi_{c}^{0/+} \pi^{+/-}$ final state~\cite{Yelton:2016fqw}. 
Belle also observed $\Xi_{c}(3055)^{+/0}$ and $\Xi_{c}(3080)^{+/0}$ in the $\Lambda D^{+/0}$ final state~\cite{Kato:2016hca}.
The ratios of branching fractions are measured as ${\cal B}(\Xi_{c}(3055)^{+} \to \Lambda D^{+})/{\cal B}(\Xi_{c}(3055)^{+} \to \Sigma_{c}(2455)^{++}K^{-})=5.09\pm1.01\pm0.76$
${\cal B}(\Xi_{c}(3080)^{+} \to \Lambda D^{+})/{\cal B}(\Xi_{c}(3080)^{+} \to \Sigma_{c}(2455)^{++}K^{-})=1.29\pm0.30\pm0.15$. These are the first measurements of
ratios of the branching fractions for (heavy-baryon and light meson) and (light-baryon and heavy meson) channels, and should be useful to identify the nature of the excited states.
Figure \ref{xics} shows the invariant mass distributions of $\Lambda D^{+}$ (left), $\Sigma_{c}(2455)^{++} K^{-}$ (middle), and $\Sigma_{c}(2520)^{++}K^{-}$ from Ref.~\cite{Kato:2016hca}.

\begin{figure*}[htbp]
  \begin{center}
    \includegraphics[scale=0.45]{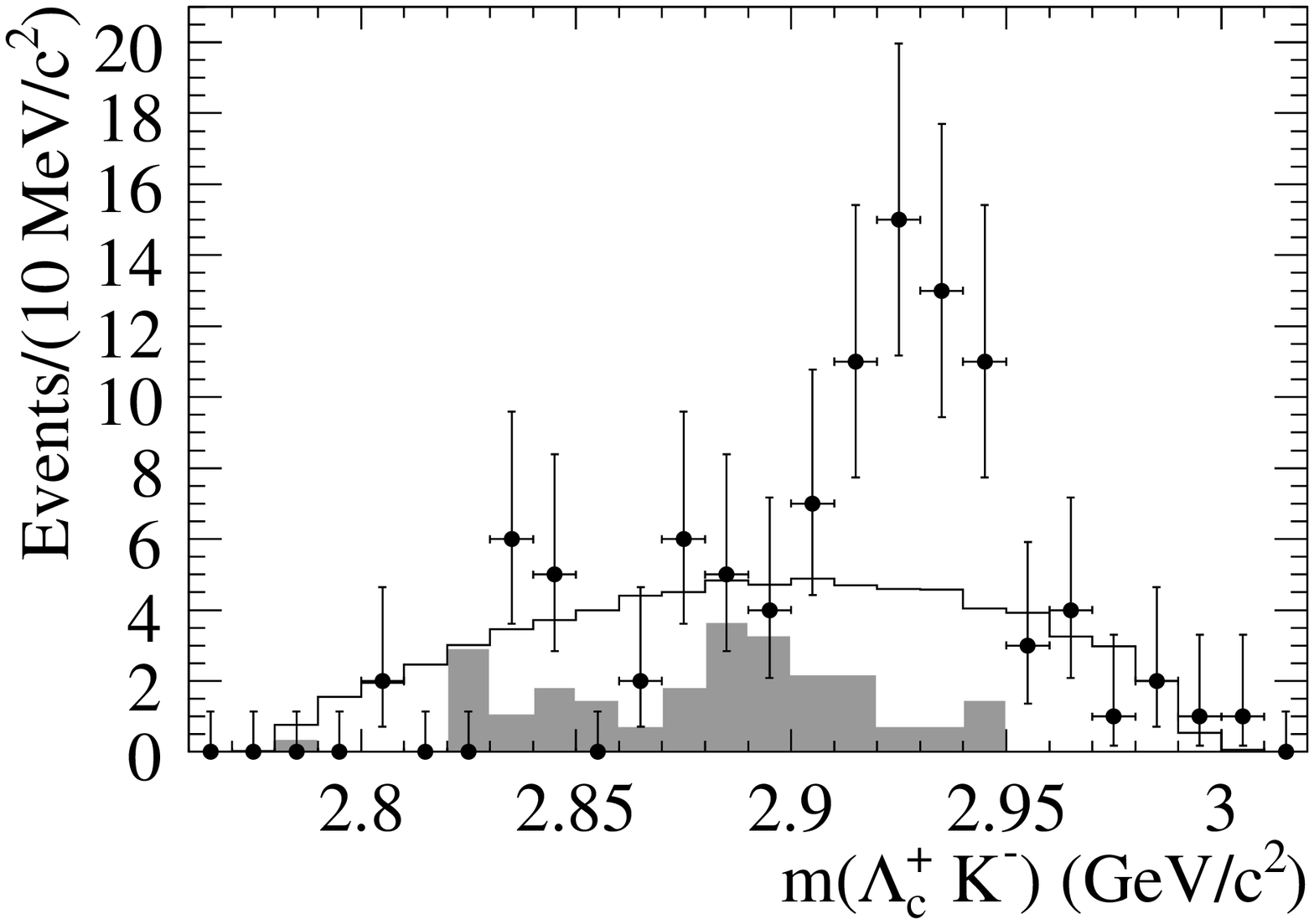}
    \includegraphics[scale=0.45]{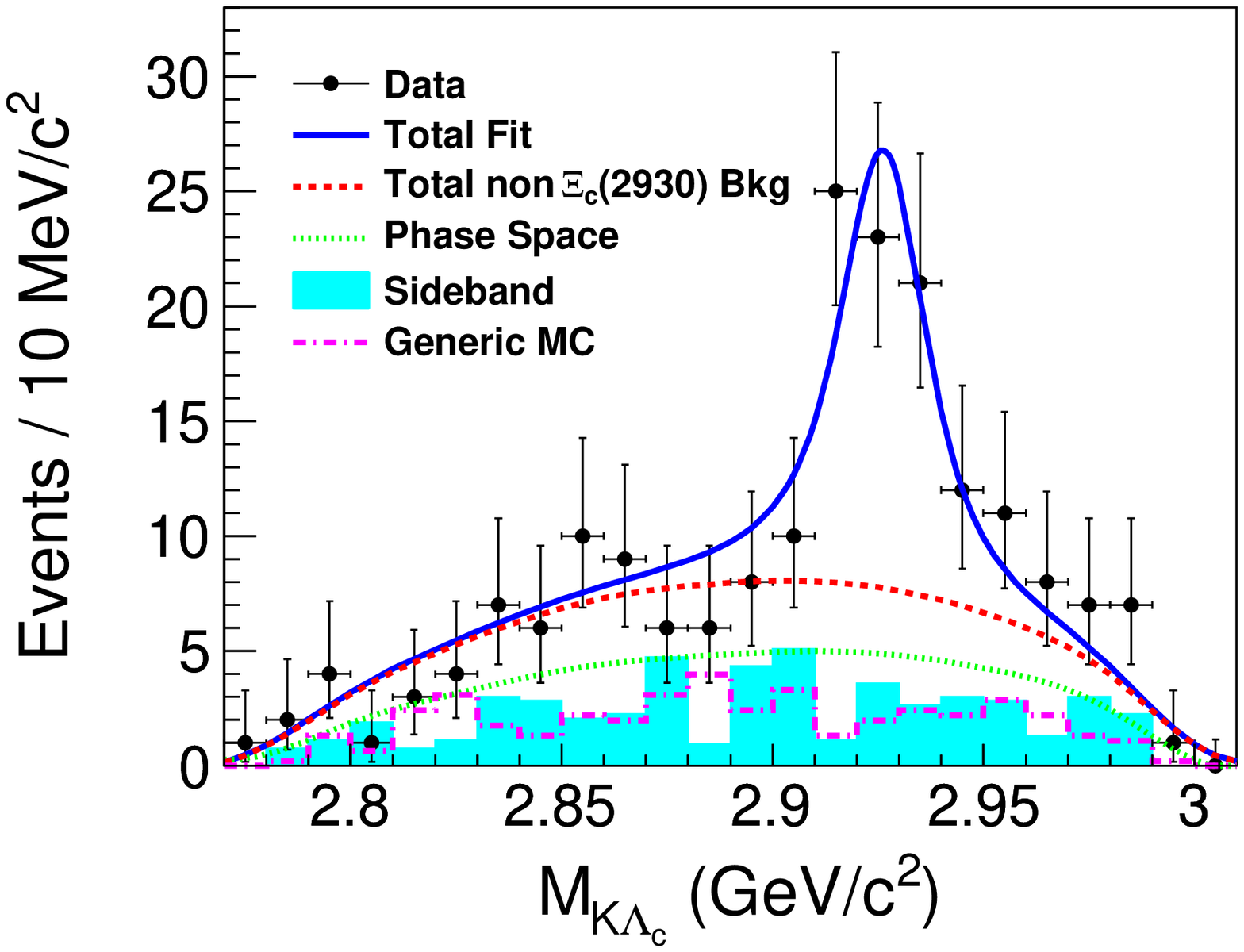}
    \caption{Invariant mass distributions of $\Lambda_{c}^{+} \bar{K}$ from BaBar~\cite{Aubert:2007eb} (left), and Belle~\cite{Li:2017uvv} (right)
             showing the evidence of $\Xi_{c}(2930)$.}
    \label{xic2930}
  \end{center}
\end{figure*}

BaBar reported a hint of a resonance, which is now called $\Xi_{c}(2930)^{0}$, decaying in to $\Lambda_{c}^{+} \bar{K}$ from the $B$ meson decay
$\bar{B} \to \Lambda_{c}^{+} \bar{\Lambda}_{c}^{-} \bar{K}$~\cite{Aubert:2007eb}, but the statistical significance not mentioned.
Later, Belle analyzed the same channel and confirmed the result~\cite{Li:2017uvv} with a significance of 5.1$\sigma$.
Figure \ref{xic2930} shows the $\Lambda_{c}^{+} \bar{K}$ invariant mass distributions, where peaks corresponding to $\Xi_{c}(2930)^{0}$ are visible.
It is worth mentioning that $\Xi_{c}(2930)$ is the first charmed baryon discovered in the $B$ meson decay.

Belle precisely measured the masses and widths for five excited charmed baryons:
$\Xi_{c}^{'}$, $\Xi_{c}(2645)$, $\Xi_{c}(2790)$, $\Xi_{c}(2815)$, and $\Xi_{c}(2970)$ using 
the final states containing $\Xi_{c}$~\cite{Yelton:2016fqw}.  
The precisions for masses relative to that of ground state $\Xi_{c}$ are improved by roughly one order of magnitude compared 
to the previous world averages.  
The first significant measurements for the widths were achieved for $\Xi_{c}(2645)^{0}$, $\Xi_{c}(2790)^{+/0}$,
and $\Xi_{c}(2815)^{+/0}$. The isospin splittings between the member of each isodoublet, which contain information
about the wave function of baryons, as written in the 
description for the $\Sigma_{c}$ family, are also measured. Table~\ref{summary_xicsplitting} summarizes the values.
Interestingly, the splittings for $\Xi_{c}(2645)$ and $\Xi_{c}^{'}$ are less then 1 MeV/c$^{2}$ whereas those for $\Xi_{c}(2790)$, $\Xi_{c}(2815)$ 
and $\Xi_{c}(2970)$ are larger than 3 MeV/c$^{2}$ with uncertainties of around 0.5 MeV/c$^{2}$. 
Qualitatively, the values are small when the Coulomb repulsive potential is small. This situation is achieved when the charge radius is large.
This naively explains the large splitting for $\Xi_{c}(2790)$ and $\Xi_{c}(2815)$, which are P-wave excited states.
These tendencies are also qualitatively consistent with the theoretical prediction~\cite{SilvestreBrac:2003kd,Guo:2008ns} except for $\Xi_{c}(2970)$ 
which was not discovered when the theoretical calculations were performed. 

\begin{center}
  \begin{table*}[htbp]
	  \caption{Summary of the isospin splittings of the $\Xi_{c}$ baryons reported in~\cite{Yelton:2016fqw}.}
     \begin{tabular}{cc}        \hline \hline
	     Particle            & $M(\Xi_{c}^{+}) - M(\Xi_{c}^{0})$ (MeV/$c^{2}$)     \\ \hline 
	     $\Xi_{c}'$          & $-0.8\pm0.1\pm0.1\pm0.5$             \\   
	     $\Xi_{c}(2645)$     & $-0.85\pm0.09\pm0.08\pm0.48$         \\   
	     $\Xi_{c}(2790)$     & $-3.3\pm0.4\pm0.1\pm0.5$             \\   
	     $\Xi_{c}(2815)$     & $-3.47\pm0.12\pm0.05\pm0.48$         \\   
	     $\Xi_{c}(2970)$     & $-4.8\pm0.1\pm0.2\pm0.5$             \\ \hline \hline
    \end{tabular}    
    \label{summary_xicsplitting}
  \end{table*}
\end{center}

Several theoretical works aimed to identify the nature of the excited states above the $\lambda$ mode P-wave excited states:
$\Xi_{c}(2930)$, $\Xi_{c}(2970)$, $\Xi_{c}(3055)$, and $\Xi_{c}(3080)$.
The masses of $\Xi_{c}(2970)$, $\Xi_{c}(3055)$, and $\Xi_{c}(3080)$ are commonly higher than those of 
$\Lambda/\Sigma_{c}(2765)^{+}$, $\Lambda_{c}(2860)^{+}$, and $\Lambda_{c}(2880)^{+}$ by around 190 MeV/$c^{2}$,
suggesting that these states are strange analogue of $\Lambda_{c}^{+}$ family.  
A theoretical calculation based on the QCD sum rules indicates $\Lambda_{c}(2860)^{+}$, $\Lambda_{c}(2880)^{+}$
and $\Xi_{c}(3055)^{+}$, $\Xi_{c}(3080)^{+}$ are two heavy quark spin doublets with a D-wave in the $\lambda$ mode
excitation~\cite{Wang:2017vtv} based on the mass of these states. This seems to be an attractive scenario.
However, other theoretical calculations indicate that
the D-wave assignment for $\Xi_{c}(3080)$ has several difficulties based on the measured branching fractions
or the total width~\cite{Liu:2012sj,Ye:2017dra,Chen:2017aqm}.
Additionally, the large decay branching fraction of $\Xi_{c}(3055) \to \Lambda D$ is not consistent with the chiral quark model 
calculation of the $3/2^{+}$ D-wave $\lambda$ mode excited state~\cite{Liu:2012sj}.
$\Xi_{c}(2970)$ is a candidate of the 2S excited state of $\Xi_{c}$~\cite{Cheng:2006dk,Ebert:2011kk}.
The large isosplitting for $\Xi_{c}(2970)$ measured by Belle should help to understand the nature of this baryon.
$\Xi_{c}(2930)$ is a candidate for the P-wave excitation state of $\Xi_{c}'$, which is possibly a 
$\Xi_{c}'$ analogue of $\Sigma_{c}(2800)$, or a candidate of the 2S excited state of
$\Xi_{c}$~\cite{Liu:2012sj,Chen:2017sci,Ye:2017yvl}.

A future determination of $J^{P}$ is important to clarify the nature of these states. 
It should be noted that no analogue state for $\Lambda_{c}(2940)^{+}$ is observed in the $\Xi_{c}$ family.
This state may possibly be observed in the future experiments such as Belle II.

\subsection{$\Omega_{c}$}
\begin{figure*}[htbp]
  \begin{center}
    \includegraphics[scale=0.4]{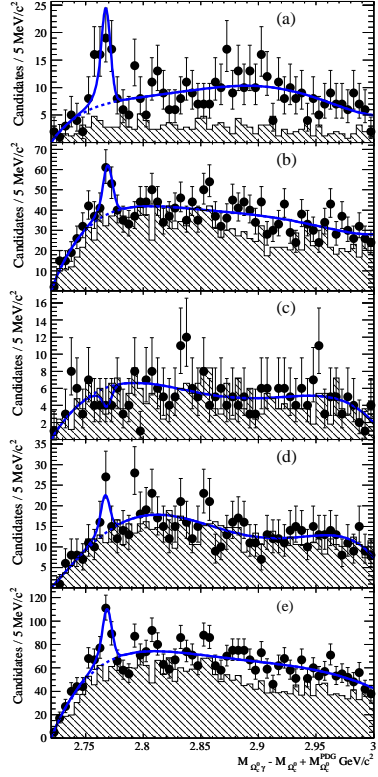}
    \caption{Invariant mass distribution of $\Omega_{c} \gamma$ for the $\Omega_{c}$ decay modes of 
             (a)$\Omega^{-} \pi^{+}$, (b) $\Omega^{-} \pi^{+} \pi^{0}$, (c) $\Omega^{-} \pi^{+} \pi^{-} \pi^{+}$, 
            (d) $\Xi^{-} K^{-} \pi^{+} \pi^{+}$, and (e) the all the decay modes combined. Figure is from~\cite{Aubert:2006je}.}
    \label{omegacstar}
  \end{center}
\end{figure*}

$\Omega_{c}$ baryons are composed of a $c$ quark and two $s$ quarks. 
Table~\ref{summary_omegac} summarizes the experimentally observed $\Omega_{c}$ family.
As the flavor part of the di-quark wave function is symmetric, the spin of di-quark must be one.
Therefore, the ground state forms a heavy quark spin doublet with total spin with $1/2$ and $3/2$.
Only the state with $1/2$, $\Omega_{c}$ was observed prior to the  B-factories.
BaBar observed the spin partner $\Omega_{c}(2770)^{0}$ in the $\Omega_{c} \gamma$ final state~\cite{Aubert:2006je}.
Figure \ref{omegacstar} shows the $\Omega_{c} \gamma$ invariant mass distributions for various $\Omega_{c}$ decay channels.
Belle confirmed BaBar's finding and is currently most precise mass measurement of $\Omega_{c}$~\cite{Solovieva:2008fw}.

LHCb observed 5 excited $\Omega_{c}$ states $\Omega_{c}(3000)^{0}$, $\Omega_{c}(3050)^{0}$,  $\Omega_{c}(3065)^{0}$, $\Omega_{c}(3090)^{0}$, 
and $\Omega_{c}(3120)^{0}$, in the $\Xi_{c}^{+} K^{-}$ final state~\cite{Aaij:2017nav} 
(The sign of $\Omega_{c}(3188)^{0}$ was also reported, but it is not significant).
Belle confirmed the existence of these states except for $\Omega_{c}(3119)^{0}$~\cite{Yelton:2017qxg}. 
Naively, five excited $\Omega_{c}$ states are expected (one $1/2^{-}$, two $3/2^{-}$, and $5/2^{-})$
in the P-wave state as the spin of the two strange quarks is one. Some of these newly discovered states should correspond to these 
P-wave states.  

\begin{center}
  \begin{table*}[htbp]
     \caption{Summary of the observed $\Omega_{c}$ family~\cite{Patrignani:2016xqp}.}
     \begin{tabular}{c|cccc}     \hline \hline
       Particle                  & Mass (MeV/c$^{2}$)                  &Width (MeV)                 & $J^{P}$   & Strong/EM decay modes   \\ \hline 
       $\Omega_{c}^{0}$          & $2695.2\pm1.7$                      & -                          & $1/2^{+}$ & -                                \\   
       $\Omega_{c}(2770)^{0}$    & $2765.9\pm2.0$                      & -                          & $3/2^{+}$ & $\Omega_{c}^{0} \gamma$          \\   
       $\Omega_{c}(3000)^{0}$    & $3000.4\pm0.2\pm0.1^{+0.3}_{-0.5}$  & $4.5\pm0.6\pm0.3$          & $?^{?}$   & $\Xi_{c}^{+} K^{-}$              \\   
       $\Omega_{c}(3050)^{0}$    & $3050.2\pm0.1\pm0.1^{+0.3}_{-0.5}$  & $0.8\pm0.2\pm0.1$ ($<1.2$) & $?^{?}$   & $\Xi_{c}^{+} K^{-}$              \\   
       $\Omega_{c}(3065)^{0}$    & $3065.6\pm0.1\pm0.3^{+0.3}_{-0.5}$  & $3.5\pm0.4\pm0.2$          & $?^{?}$   & $\Xi_{c}^{+} K^{-}$              \\   
       $\Omega_{c}(3090)^{0}$    & $3090.2\pm0.3\pm0.9^{+0.3}_{-0.5}$  & $8.7\pm1.0\pm0.8$          & $?^{?}$   & $\Xi_{c}^{+} K^{-}$              \\   
       ($\Omega_{c}(3120)^{0}$)  & $3119.1\pm0.3\pm0.9^{+0.3}_{-0.5}$  & $1.1\pm0.8\pm0.4$ ($<2.6$) & $?^{?}$   & $\Xi_{c}^{+} K^{-}$              \\ \hline
    \end{tabular}    
    \label{summary_omegac}
  \end{table*}
\end{center}

\section{Prospect at the SuperKEB/Belle II experiment}
As shown in previous sections, B-factory experiments have played a significant role in the field of open charm hadron spectroscopy
in the last 15 years. However, the nature of the observed hadrons remains uncertain. Moreover, many states  
have been predicted but have yet to be observed.
More research experimentally and theoretically is necessary to achieve a comprehensive understanding of open charm hadrons. \par
SuperKEKB/Belle II is a upgrade of the KEKB/Belle, which aims to collect a data set with 50 times
integrated luminosity to search for the physics beyond the Standard Model. 
The KEKB accelerator has been upgraded to SuperKEKB, which is designed to have a peak luminosity 40 times
higher than that of KEKB~\cite{Ohnishi:2013fma}. This is achieved by utilizing so called the ``nano-beam scheme",
where the beam size in the vertical axis is squeezed in around 50 nm at the interaction point~\cite{Bona:2007qt},
and doubling the beam current.
The Belle detector has also been upgraded to the Belle II detector to cope with expected high beam originated background
 in SuperKEKB and to improve the performance~\cite{Abe:2010gxa}.
At the time of this writing, Belle II experiment has just started the operation but only part of the vertex detector installed.
Well-known hadrons such as $D^{(\ast)}$ or $B$ have already been rediscovered.
The operation with a full detector setup is scheduled for early 2019.  \par
SuperKEKB/Belle II will allow us to study not only the physics beyond the Standard Model,
but also hadron spectroscopy more deeply. For example, it is experimentally known that the hadron production 
cross section in the $e^{+}e^{-}$ collision drops in the exponential curve as a function of hadron mass~\cite{Niiyama:2017wpp}.
This means that the current B-factory data may miss a high mass resonance due to the small cross sections.
Belle II will make it possible to discover these hadrons.
The determination of the quantum number is very important to identify the nature of excited states.
As shown in the example of the determination of the spin of $\Sigma_{c}(2455)$~\cite{Aubert:2008ax}, 
the charmed hadrons produced from the $B$ meson is very useful to determine their quantum numbers
from the constraints on the helicities. Collecting more data should allow to observe higher 
spin state in the $B$ decays and spin to be determined in a model independent way.
Additionally, several theoretical papers have indicated that measuring various possible decays is 
important to understand the nature of hadrons.
Many decay modes have yet to be surveyed such as those including $\eta$ mesons. 
It is important to study these decay modes comprehensively with a high statistics data sample.

\section{Summary}
In summary, there has been a significant progress in the field of open charm hadron spectroscopy,
which has been driven by B-factory experiments, Belle and BaBar.
More than fifteen new charmed mesons and baryons are discovered in this fifteen years
and their properties are measured in a comprehensive manner.
In the $D$ meson family, two states which can be interpreted as $L=1$ and $j_{q}=1/2$ are discovered.
In the $D_{S}$ familiy, two interesting states $D_{s 0}^{\ast} (2317)^{+}$ and $D_{s 1}(2460)^{+}$ are observed.
Naively, they are too light to be interpreted as ordinal $c\bar{s}$ states with  $L=1$ and $j_{q}=1/2$. 
Many higher excited states are observed in the continuum productions for both $D$ and $D_{S}$ states and 
$J^{P}$ are determined for some of them. 
In the baryon sector, new excited states are observed in $\Lambda_{c}$, $\Sigma_{c}$, $\Xi_{c}$, and $\Omega_{c}$ families.
With the discovery of $\Omega_{c}(2770)$, all the ground states predicted by quark model appeared all together.
Newly discovered higher excited states are expected to be D-wave excited states, radial excited states, and possibly meson-baryon molecular states.
To identify the nature, determination of $J^{P}$ is essential. However, $J^{P}$ of these states are almost not determined except for $\Lambda_{c}(2880)^{+}$ by Belle 
and $\Lambda_{c}(2940)^{+}$ by LHCb.
There are still a lot of open questions in the open charm hadrons. Hope these will be answered by the Belle II experiment which has just observed its first collision.

\section{Acknowledgements}
This work is supported by Grant-in-Aid for Scientific Research (S) (No.26220706) and
 Grant-in-Aid for Scientific Research on Innovative Areas ``Elucidation of New Hadrons with a Variety of Flavors''.

\bibliography{refs}

\end{document}